\documentclass[10pt,journal,compsoc]{IEEEtran}
\usepackage[T1]{fontenc}% optional T1 font encoding
\usepackage[english]{babel}
\usepackage{graphicx,caption}
\usepackage{subfig}
\usepackage{amsthm}
\usepackage{float}
\usepackage{epsfig}
\usepackage{url}
\usepackage{multirow}
\usepackage{graphicx}
\usepackage{amsmath}
\usepackage{amssymb}
\usepackage{mathrsfs}
\usepackage{amsfonts}
\usepackage{paralist}
\usepackage{tabularx}
\usepackage{booktabs}
\usepackage{color,xcolor}

\usepackage{algorithm}
\usepackage{algpseudocode}
\theoremstyle{remark}

\theoremstyle{definition}

\newfloat{problem}{htb}{loM}
\floatname{problem}{Problem}

\ifCLASSOPTIONcompsoc
  \usepackage[nocompress]{cite}
\else
  \usepackage{cite}
\fi
\ifCLASSINFOpdf
\else
\fi

\begin{document}
\graphicspath{{images/}}
\title{When Deep Reinforcement Learning Meets Federated Learning: Intelligent Multi-Timescale Resource Management for Multi-access Edge Computing in 5G Ultra Dense Network}
\author{Shuai~Yu, Xu~Chen, Zhi~Zhou, Xiaowen~Gong, and Di~Wu

\thanks{Shuai~Yu, Xu~Chen, Zhi~Zhou and Di~Wu are with the School of Data and Computer Science, Sun Yat-sen University, Guangzhou, China, (E-mail: \{yushuai, chenxu35, zhouzhi9 and wudi27\}@mail.sysu.edu.cn). }
\thanks{Xiaowen Gong is with the Department of Electrical and Computer Engineering at Auburn University, Auburn, AL, USA, (E-mail: xgong@auburn.edu). }
\thanks{Corresponding Author: Xu Chen.}
}
\IEEEtitleabstractindextext{%
\begin{abstract}
Recently, smart cities, healthcare system and smart vehicles have raised challenges on the capability and connectivity of state of the art Internet of Things (IoT) devices, especially for the devices in hot spots area.
Multi-access edge computing (MEC) can enhance the ability of emerging resource-intensive IoT applications and has attracted much attention.
However, due to the time-varying network environments, as well as the heterogeneous resources of network devices, it is hard to achieve stable, reliable and real-time interactions between edge devices and their serving edge servers, especially in the 5G ultra dense network (UDN) scenarios.
Ultra-dense edge computing (UDEC) has the potential to fill this gap, especially in the 5G era, but it still faces challenges in its current solutions, such as the lack of:
i) efficient utilization of multiple 5G resources (e.g., computation, communication, storage and service resources);
ii) low overhead offloading decision making and resource allocation strategies;
and iii) privacy and security protection schemes.
Thus, we first propose an intelligent ultra-dense edge computing (I-UDEC) framework, which integrates blockchain and Artificial Intelligence (AI) into 5G ultra-dense edge computing networks.
%First, we show the architecture of the framework.
Then, in order to achieve real-time and low overhead computation offloading decisions and resource allocation strategies, we design a novel two-timescale deep reinforcement learning (\textit{2Ts-DRL}) approach, consisting of a fast-timescale and a slow-timescale learning process, respectively.
The primary objective is to minimize the total offloading delay and network resource usage by jointly optimizing computation offloading, resource allocation and service caching placement.
We also leverage federated learning (FL) to train the \textit{2Ts-DRL} model in a distributed manner, aiming to protect the edge devices' data privacy.
Simulation results corroborate the effectiveness of both the \textit{2Ts-DRL} and FL in the I-UDEC framework and prove that our proposed algorithm can reduce task execution time up to $31.87\%$.

\end{abstract}

}

\maketitle

\IEEEdisplaynontitleabstractindextext

% \fi
\IEEEpeerreviewmaketitle

\IEEEraisesectionheading{\section{Introduction}\label{sec:introduction}}

\IEEEPARstart{N}owadays, the advances in cloud computing and communication technology are spawning a variety of intelligent mobile and IoT devices, such as smart phones, Google Glass and smart meters.
Accompanied by the intelligent devices, multiple resource-intensive mobile and IoT applications are designed to enhance the user experience, such as wearable cognitive assistance, augmented reality, and collaborative 3D gaming.
Such applications use complex algorithms for camera tracking and object recognition, require extensive computation power, high speed connection, real-time feedback and a variety of cloud computing services.
Furthermore, the number of such mobile and IoT devices is anticipated to have an explosive growth in the 5G era, which will result in the congestion of the cloud computing network.
In light of this, multi-access edge computing (MEC)~\cite{ETSIMEC,9061048} and ultra dense network (UDN)~\cite{UDN} are expected as effective solutions to meet the computation and communication requirements, respectively.
MEC provides a capillary distribution of cloud computing capabilities to the edge of network, enabling rich services and applications in close proximity to edge devices.
UDN can offer adequate base band resources by deploying more base stations close to the edge users.

Recently, a new research paradigm named ultra-dense edge computing (UDEC)~\cite{8436039} emerges.
It integrates MEC with UDN by deploying MEC servers on the base stations of UDN~\cite{8647895,8058414}.
This integration not only achieves lower data transmission delay, but also meets the requirement of real-time computing.
%MEC could be deployed at the edge of SD-UDN, and is able to provide a variety of real-time services to edge users by computation offloading~\cite{CoarseOffloadingZhang, FineOffloadingKao} and edge service caching~\cite{service_caching}.
%Computation offloading consists of application partitioning and resource allocation~\cite{7879258}, plays a critical role in MEC networks.
%Service caching refers to caching application services and their related databases/libraries in the MEC servers, thereby enabling corresponding computation tasks to be executed closer to edge users.
However, how to effectively integrate MEC with UDN becomes a critical problem in the future development of 5G, and is under-explored to the best of our knowledge.
Note that, an effective integration should consider the following key issues:
\textbf{i) Full use of system resources:}
by leveraging computation offloading~\cite{8436039}, a mobile application can be split into local parts (usually communication intensive) and offloaded parts (usually computation intensive).
Then, an edge server allocates a variety of system resources (e.g., computation, communication, storage and service) to process the offloaded parts~\cite{8463562}.
The problem becomes extremely complex in the UDEC environments, since UDEC has massive edge devices and edge servers with heterogeneous computation resources (e.g., CPU, GPU) and communication links (e.g., cellular and D2D).
Thus, it is necessary to fully and efficiently utilize the system resources to improve the UDEC performance.
\textbf{ii) Dynamic and low-cost scheduling:}
the system resources in UDEC are usually diverse (e.g., various service caching placement strategies, different CPU frequencies of mobile devices and cloud/edge servers, memory) and time-varying (e.g., network disconnection, channel state and backhaul latency).
The distributed and heterogeneous natures of the resources make the scheduling process more complex.
Moreover, computation offloading requires real-time application partitioning and low overhead resource allocation.
Thus, dynamic and real-time scheduling schemes are required to enhance the quality of service (QoS) of edge devices.

\textbf{iii) Privacy\&security-preserving services:} are crucial to the MEC-enabled UDN, since the interplay of heterogeneous edge devices and the migration of services across devices cause a lot of security and privacy issues.
Specifically, for many cloud services, their related databases/libraries usually contain edge users' sensitive information (e.g., location, account information and medical records in real-life situations).
Edge users may not trust the remote cloud server, thus unwilling to offload the computation with their private data to the server.
Therefore, edge users are prone to offload their private data to their trusted nearby edge servers with privacy-preserving techniques, such as Points-of-Interest Services and Traffic Information Services~\cite{8737758}.

In this context, an integrated framework is required to jointly optimizes the application partitioning, resource allocation and service caching placement for UDEC with privacy protection.
However, the delay sensitivity varies for the above issues, and the complexity of the optimization problem is very high for the joint optimization problem.
Recently, deep learning~\cite{9062302} and blockchain~\cite{8624417} have attracted much attention of researchers in edge computing fields.
Deep learning is widely exploited in wireless communication networks (e.g., 4G and VANET) to obtain timely decision making (e.g., automatic resource allocation and robotics).
Blockchain provides reliable connections and management of the computation, communication and storage resources within the massive distributed edge nodes of UDEC~\cite{8624417}.
Thus, there is a pressing need to explore a novel deep learning and blockchain based scheduling approach for UDEC.
In addition, federated learning (FL)~\cite{DBLP:journals/corr/KonecnyMYRSB16,9090366,8770530} is regarded as a useful tool for training deep learning agents in a distributed manner, as FL can protect the personal data privacy in the model training process.

In this article, we propose an intelligent ultra-dense edge computing (I-UDEC) framework that jointly optimizes the issues of application partitioning, resource allocation and service caching placement in ultra-dense edge computing environments.

The major contributions of this article are summarised as follows:
\begin{itemize}
\item First, we propose an intelligent ultra-dense edge computing (I-UDEC) framework and formulate the heterogeneous resources (including computation, communication and service) of the I-UDEC.
    We also introduce a hybrid computation offloading strategy for the I-UDEC, in which an edge user can offload its resource-intensive tasks to: i) a remote cloud server (i.e., cloud computing), ii) a nearby edge server (i.e., edge computing), or, iii) a nearby mobile device through device-to-device (D2D) computation offloading.
%\item Then, we propose a comprehensive privacy protection mechanism which consists of: i) social trust-based D2D offloading in user plane, ii) data sensitivity-based computation offloading in data plane, and iii) federated learning-based model training in control plane for the I-UDEC.
\item Then, since application partitioning, resource allocation and service caching placement have different delay sensitivities, we present a two-timescale deep reinforcement learning (\textit{2Ts-DRL}) approach to jointly optimize the above issues for the I-UDEC.
    Specifically, the \textit{2Ts-DRL} consists of two tiers and thus operates in two different timescales.
    The bottom tier of the \textit{2Ts-DRL} outputs delay sensitive decisions (i.e., application partitioning and resource allocation strategy) in a fast timescale, whereas the top tier outputs delay insensitive decision (i.e., service caching placement strategy) in a slow timescale.
\item Last but not least, we use a FL-based distributed model training method to train the DRL agent, in order to protect edge users' sensitive service request information.
%DRL is the state-of-the-art reinforcement learning approach that approximate the Q value-action function by leveraging deep Q network (DQN)~\cite{8061008}.
\end{itemize}

The rest of this article is organized as follows.
Section~\ref{sec:RelatedWork} presents an overview of the related works.
In Section~\ref{sec:I-UDEC}, the architecture, system model and computation offloading process are presented for the proposed intelligent ultra-dense edge computing (I-UDEC) framework.
Section~\ref{sec:Problem_Formulation} formulates the optimization problems, and introduces our proposed \textit{2Ts-DRL} algorithm.
A federated learning-based distributed model training method is presented in Section~\ref{sec:Federated_Learning}.
Simulation results are discussed in Section~\ref{sec:performance_evaluation}.
Finally, conclusions are presented in Section~\ref{sec:conclusion}.

\section{Related Work}{\label{sec:RelatedWork}}

\subsection{Ultra-dense Edge Computing}
To integrate cloud computing functionalities into wireless networks, European Telecommunications Standards Institute (ETSI) develops the concept of multi-access edge computing (MEC)~\cite{ETSIMEC} that is expected to play a key role in meeting the requirements of 5G.
Edge servers, which are located at the edge of MEC networks, can provide computing resource, storage capability, connectivity and various services for edge devices.
However, due to the limited computation resource and storage capability of the edge servers, only a few edge devices can be served by each single server.
Thus, dense deployment of edge servers in MEC is required, especially in some heavily populated urban areas (e.g., shopping malls and train stations).
Ultra-dense edge computing (UDEC)~\cite{8436039} is designed to tackle the above challenges, by deploying MEC servers on the base stations in ultra dense network (UDN).
%The new edge servers are called small cell cloud-enhanced e-Node B (SCceNB)~\cite{TROPIC}.
Due to the dense deployment of the edge servers, UDEC can provide a huge computing functionalities and a variety of services close to edge users, in order to meet the low latency demand of 5G.
For example, authors in~\cite{8058414} study the mobility management issue of UDEC.
They propose a new user-centric energy-aware mobility management (EMM) scheme to optimize communication and computation delay.
At the same time, they also consider the long-term energy consumption constraints.

Recently, a lot of works pay much attention to the computation offloading issue of UDEC.
For example, in order to reduce energy consumption, authors in~\cite{8436039} propose a greedy-based offloading scheme in a multi-user ultra-dense MEC server scenario.
However, only one mobile device is considered in their model.
Authors in~\cite{8647895} study the computation offloading issue in UDEC.
They propose a game-theoretical offloading scheme that can jointly optimized task offloading, computation frequency scaling and transmit power allocation.
Thus, the energy consumption on the mobile terminal side can be minimized.
However, service caching placement is ignored in this article.

\subsection{Artificial Intelligence in Edge Computing}
Recently, edge intelligence (EI)~\cite{8976180,8876870,9127160,9094236,8270633,9151375}, aiming to integrate artificial intelligence (especially deep learning) services into edge computing, has attracted much attention.
On the one hand, EI aims to push artificial intelligence computations from remote cloud to the network edge, thus to achieve various real-time and reliable intelligent services.
For example, authors in~\cite{7837725} design a food recognition system for dietary assessment by leveraging EI.
The system operates in the edge computing environments, aims to obtain the best-in-class recognition accuracy, as well as real-time recognition.
On the other hand, EI also has the potential to solve various network optimization problems (i.e., resource allocation and offloading decision).
For example, authors in~\cite{8061008} present an integrated framework that jointly optimize resource allocation, content caching, and computation offloading for vehicular network.
In order to solve the problem of high complexity of the joint optimization problem, they propose a deep reinforcement learning (DRL) based approach to achieve automatic decision making.
However, the corresponding training time is very long, the delay sensitivity for different optimization objectives (e.g., wireless resource allocation and content caching placement) is neglected.

\begin{table}[t]
\caption{Key Notation}
\label{table:keynotations}
 \begin{tabular}{m{1in}<{}m{2.1in}<{}}\hline
 \toprule
\textbf{Symbol}&\textbf{Definition}\\
  \midrule
ED  & edge device, including the 5G mobile users and IoT devices.\\
SCceNB &edge server, refers to the small cell cloud-enhanced e-Node B. \\
$\mathcal{M} = \{1, 2,..., M\}$ & The set of EDs. \\
$\mathcal{N} = \{1, 2,..., N\}$ & The set of SCceNBs. \\
$C_{s}$ &The storage capacity for the SCceNBs.\\
$f_{s}$, $f_{u}$ &The CPU frequency for each single CPU core of the SCceNBs and EDs, respectively.\\
$\boldsymbol{Q}_{SC}$, $\boldsymbol{Q}_{ED}$ & The service caching placement index for the SCceNBs and EDs, respectively.\\
$\mathcal{O}=(\mathcal{V},\mathcal{D})$ & Task model, where $\mathcal{V}$ represents the set of sub-tasks, $\mathcal{D}$ denotes the data dependencies between the sub-tasks. \\
$\xi_{v}$                             & The workload (CPU cycles/number of instructions to be executed) of subtask $v$.\\
$\rho_{v}$                            & Required cpu cycles a CPU core will perform per byte for the input data processed by the sub-task $v$.\\
$\boldsymbol G_{n}(t)$                & The task queue phase of SCceNB $n$ in decision period $t$. \\
$\boldsymbol{E}_{n}(t)$               & Offloading action for the task queue $\boldsymbol G_{n}(t)$ in decision period $t$. \\
$\boldsymbol S_{n}(t)$                & The composite system state of SCceNB $n$ at each decision period $t$.\\
$\boldsymbol K_{n}(t)$                & Wireless resource (i.e., subcarrier) allocation strategy in decision period $t$. \\
  \bottomrule
 \end{tabular}
\end{table}

\subsection{Blockchain in Edge Computing}
Thanks to the advantages of security, privacy and scalability of blockchain.
Integrating blockchain to edge computing has been widely studied in recent years.
First, blockchain plays a key role in the secure data transmission for edge computing networks.
For example, the authors in~\cite{8053750} design a blockchain-based distributed cloud architecture to provides low-cost and secure access to the fog computing networks.
The proposed framework also uses software defined networking (SDN) based controller to minimized the end-to-end delay between edge devices and system resources.
Then, the decentralized P2P data storage manner of blockchain can increase the availability and scalability of edge computing networks.
Since edge computing networks keep the data close to edge devices in separate storage locations.
For example, authors in~\cite{8884875} propose a novel blockchain system for edge computing networks.
The system stores metadata items onto blocks instead of actual data of end users by compressing the storage of blocks.
They also provide a strategy that can achieve optimal places to store data and blocks.
Last but not least, blockchain also enables computation offloading in edge computing networks.
Authors in~\cite{8884875} propose a blockchain-enabled computation offloading strategy in mobile edge computing environments.
The main objective of the strategy is to guarantee data integrity in the computation offloading process.
They also use the nondominated sorting genetic algorithm III (NSGA-III) to achieve a balanced resource allocation strategy.

In this article, we present a novel two-timescale deep reinforcement learning (\textit{2Ts-DRL}) algorithm.
The algorithm consists of a fast-timescale and a slow-timescale learning process, in order to achieve real-time and low overhead computation offloading decisions and resource allocation strategies in a fast-timescale and optimized service caching placement scheme in a slow-timescale.
We also leverage federated learning (FL) to train the \textit{2Ts-DRL} model in a distributed manner, aiming to protect the edge devices' data privacy and reduce the training overhead.
By this way, the total offloading delay and network resource usage can be optimized.

\section{Proposed intelligent ultra-dense edge computing}
\begin{figure}[t!]
    \centering
    \includegraphics[width=3.3in]{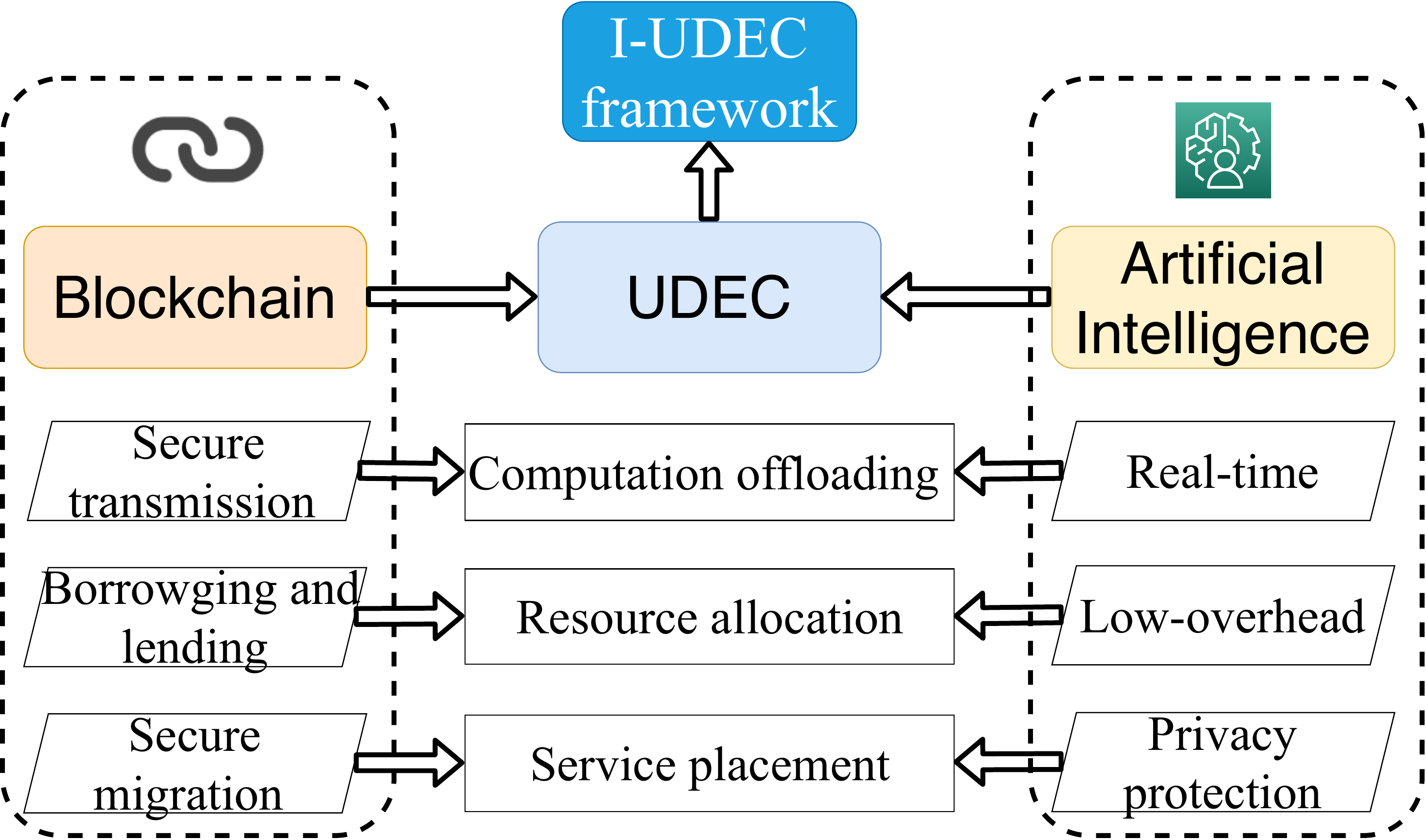}
    \caption{Artificial intelligence and blockchain enhanced ultra-dense edge computing framework.}
    \label{fig:AI-Blockchain}
\end{figure}

{\label{sec:I-UDEC}}
In this section, we will present an artificial intelligence and blockchain enhanced ultra-dense edge computing framework, i.e., the intelligent ultra-dense edge computing (I-UDEC), as shown in Fig.~\ref{fig:AI-Blockchain}.
Specifically, we first present the architecture of the I-UDEC framework.
Then, the network model and service caching model for the I-UDEC will be presented.
At last, we elaborate the computation offloading process for I-UDEC.
For ease of reference, the key notations for system model are shown in TABLE.~\ref{table:keynotations}.

\subsection{The Architecture of I-UDEC}
The I-UDEC is realized by enhancing the following three main functions of UDEC, i.e., i) computation offloading, ii) resource allocation and iii) service caching placement.
Computation offloading plays a critical role of UDEC, since edge devices can offload their computation intensive tasks to their nearby resourceful edge servers.
Thus, real-time offloading decision making can speed up the task execution time as well as saving the battery lifetime of edge devices.
To this end, the computation offloading process is enhanced by an artificial intelligence approach (i.e., deep Q-learning, as will be explained in Section~\ref{sec:Problem_Formulation}) to achieve real-time decision making.
On the other hand, during computation offloading, some security issues (e.g., jamming attacks and sniffer attacks) could be launched to disable the UDEC.
Thus, secure blockchain communication (including  encryption, addressing and so on) is applied to ensure the efficient and reliable cooperation of UDEC networks.
In UDEC, computation, communication and storage resources are distributed at the network edge, thus are hard to be managed.
To this end, an AI-based automatic resource allocation scheme is designed (as will be explained in Section~\ref{sec:Problem_Formulation}) for the I-UDEC.
Besides, a blockchain based dynamic coordination mechanism (involves resource borrowing and lending) should be considered for the edge servers.
At last, a federated learning based model training method is presented to protect the edge devices' data privacy.
Blockchain-based service migration is considered to guarantee the security during the migration of services across multiple edge servers.

\begin{figure*}[t!]
    \centering
    \includegraphics[width=6in]{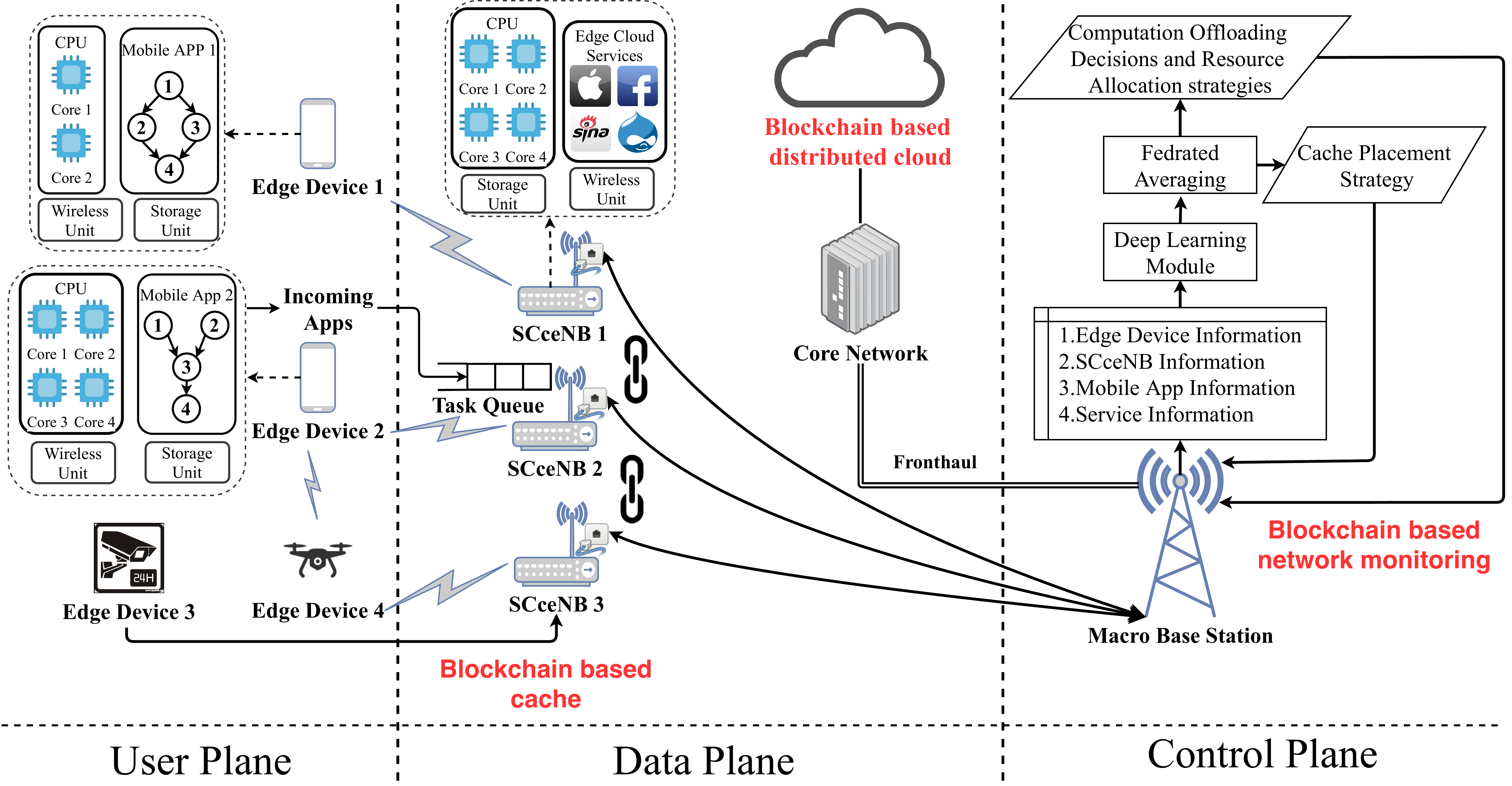}
    \caption{Illustration of intelligent ultra-dense edge computing (I-UDEC) framework for ultra-dense networks.}
    \label{fig:I-UDEC}
\end{figure*}

Specifically, the proposed I-UDEC framework is illustrated in Fig.~\ref{fig:I-UDEC}.
It consists of: i) user plane, ii) data plane and iii) control plane.
The user plane is composed of the edge devices (EDs, such as 5G smart phones and IoT sensors) who have computation offloading requirements, as shown in Fig.~\ref{fig:I-UDEC}.
The data plane corresponds to i) remote cloud server, and ii) edge servers deployed close to EDs.
In addition, blockchain-based data plane cache is also applied to i) solve the problem of saturation attacks by caching the missing packets, or ii) deal with the flooding attacks by caching the flood packages.
The control plane is realized by UDN controller deployed at macro base station (MBS).
The control functionalities (i.e., monitoring, computation offloading actions, resource allocation strategies and service caching placement) are centralized at the controller (i.e., the macro base station in Fig.~\ref{fig:I-UDEC}).
Note that blockchain-based network monitoring~\cite{8624417} is applied in this plane to prevent malicious behaviors (e.g., DDoS attack, packet saturating).
From a global view of the system states, the controller is responsible for collecting and maintaining the lists of EDs' information, edge server information, service information and application information, so as to optimize the network configurations on demand.
The ED information list includes data such as the computation and communication capacities for each ED.
edge server information list maintains the data of computation, storage and communication capacities, as well as the fronthaul delay for each edge server.
Service information list consists of current service caching placement strategy for the edge servers and EDs.
Application information includes computation offloading requirements, which consists of the application type, required service type, data amount and required CPU cycles.
Typically, an ED reports the edge device information and application information to the nearest serving edge server, when it has computation offloading requirements.
Then, the edge server integrates multiple EDs' information and edge server information together and transmits to the controller.
The controller receives and trains the data through deep learning and federated learning modules (as will be explained in Section~\ref{sec:Federated_Learning}).
At last, the controller makes i) joint computation offloading and resource allocation decisions for the edge servers and EDs in a fast timescale and ii) service caching placement strategies for the edge servers in a slow timescale.

\subsection{Network Model}
%Nowadays, mobile and IoT devices adopt the multi-core CPU architecture \cite{multicore} to increase the processing speed and extend battery life.
%Based on this CPU architecture, the edge devices in I-UDEC can process multiple tasks at the same time.
In this article, we consider the small cell-based MEC \cite{TROPIC,FemtoTMC}.
The edge servers of the small cell-based MEC are called small cell cloud-enhanced e-Node B (SCceNB).
The EDs communicates to the SCceNBs through wireless link, while the SCceNBs are connected to the MBS and remote cloud server via high speed fronthaul links.
Compared with the edge devices, SCceNBs are equipped with more computation and storage resources.
We consider our I-UDEC framework operates in an ultra-dense edge computing network that formed by a remote cloud server, a macro base station, a set $\mathcal{N} = \{1, 2,..., N\}$  of SCceNBs and a set $\mathcal{M} = \{1, 2,..., M\}$ of EDs within the coverage of SCceNB $n$ $(n \in\mathcal{N})$, as shown in Fig.~\ref{fig:I-UDEC}.
%Assume that the ED $m$ $(m\in\mathcal{M})$ is equipped with $c_{m}$ cpu cores, and each SCceNB is equipped with $c_{s}$ cpu cores.
Assume that the ED $m$ $(m\in\mathcal{M})$ is equipped with $X_{m}$ cpu cores, each SCceNB is equipped with $Y$ cpu cores.
Thus, each SCceNB can server at most $Y$ EDs at the same time.
%Due to the limited computation and storage resources, we assume that each SCceNB can server at most $M$ EDs that is denoted by $\mathcal{M} = \{1, 2,..., M\}$.
%Similarly, the number of CPU cores for different SCceNB is randomly chosen from the set $\{1,2,4,8,16\}$.
Let $C_{s}$ denotes the storage capacities for SCceNB $n$.
$f_{s}$ and $f_{u}$ represent the CPU frequency for each single CPU core of the SCceNBs and EDs, respectively.

In our I-UDEC framework, each ED can offload its tasks through establishing a cellular link with a SCceNB, or a D2D link with another ED.
Assume that the cellular and D2D links are based on the orthogonal frequency division multiple-access (OFDMA)~\cite{OFDMA}.
It means that the $M$ EDs in the coverage of SCceNB $n$ are separated in the frequency domain, and thus do not interfere with each other.
For the sake of simplicity, we utilize a subcarrier-based fixed channel assignment (FCA) scheme~\cite{FCA} within the $M$ EDs.
There are $K$ available subcarriers for a SCceNB to serve EDs.
%$\mathcal{K}=\{1,2,...,K\}$ denote the available subcarriers to be allocated.
The bandwidth of each subcarrier is $B$.

\subsubsection{Cellular Link Model}
%As stated earlier, each mobile device can establish a cellular link with a SCceNB.
During one scheduling period, assume that the locations of EDs are fixed and the wireless channels between EDs and SCceNSs are stable.
Note that they may change in different scheduling periods due to EDs' mobility.
Based on the above assumptions, the maximum uplink data rate (in bps) between ED $m$ $(m\in \mathcal{M})$ and SCceNB $n$ $(n\in \mathcal{N})$, over an additive white Gaussian noise (AWGN) channel, can be expressed as follows:
% \cite{dlulrate1, dlulrate2}
\begin{equation}\label{eq:channel}
r_{m,n}^{ul}=K_{m}^{n}\cdot B\cdot \log_{2}\left(1+\frac{p_{m}^{u}|h_{m,n}^{ul}|^{2}}{\Gamma(g_{ul})d_{m,n}^{\beta}N_{0}}\right),
\end{equation}
where $p_{m}^{u}$ refers to the transmit power for the ED $m$, $d_{m,n}$ denotes the distance between ED $m$ and SCceNB $n$, and $N_{0}$ denotes the noise power.
$K_{m}^{n}$ represents that the SCceNB $n$ allocates $K_{m}^{n}$ subcarriers to ED $m$ (i.e., a communication resource allocation scheme).
In the Rayleigh-fading environment, $h_{m,n}^{ul}$ denotes the channel fading coefficient between ED $m$ and SCceNB $n$, $g_{ul}$ refers to the target uplink BER and $\beta$ is the path loss exponent.
$\Gamma(BER)=-\frac{2log(5BER)}{3}$ is the SNR margin introduced to meet the desired target bit error rate (BER) with a QAM constellation.
In this work, we assume that these parameters are uncontrollable, similar to the assumptions that are made in~\cite{DREAM}.
Then, the uplink rate index is fed to an uplink rate weight matrix $\boldsymbol R_{n}^{UL}=[r_{m,n}^{ul}]_{1\times M}$.
Note that in the ultra-dense edge computing network, the EDs reserve a part of resource blocks for transmitting pilot signals.
The SCceNB can estimate the uplink channels by comparing the received signals and the pilot signals.
Finally, the SCceNBs send the estimated channels to the controller through via high speed fronthaul links.
In practice, the channel gain can be well estimated (with small random noise).
If the channel state is not perfect estimated, our proposed deep reinforcement learning can actually achieve a very good approximation of the true state due to the powerful learning capability of deep learning over rich experienced learning samples, which can be verified by numerical successful application examples of deep reinforcement learning including intelligent wireless communications~\cite{8594714,8917869}.

\subsubsection{D2D Link Model}
In the proposed I-UDEC framework, each ED can offload his computation to another ED in proximity through a D2D link.
For example, two edge devices ED 4 and ED 2 (as shown in Fig. \ref{fig:I-UDEC}) can offload computation to each other if their distance is less than a pre-defined threshold $R^{d}$, as in~\cite{D2D_R}.
Let $r_{i, j}^{d2d}$ denote the D2D data rate from ED $i$ to $j$ as follows:
\begin{equation}{\label{eq:downlinkrate}}
r_{i, j}^{d2d}=K_{m}^{n}\cdot B\log_{2}\left(1+\frac{p_{i}^{d2d}|h_{i, j}^{d2d}|^{2}}{\Gamma(g_{d2d})d_{i, j}^{\beta}N_{0}}\right), (i, j\in \mathcal{M}),
\end{equation}
where $h_{i, j}^{d2d}$ is the channel fading coefficient for D2D link, $p_{i}^{d2d}$ refers to the D2D transmission power of ED $i$, $d_{i, j}$ represents the distance between ED $i$ and $j$ and $g_{d2d}$ denotes the target BER for D2D link.
Then, the D2D rate index is fed to an D2D rate weight matrix $\boldsymbol R_{n}^{D2D}=[r_{i, j}^{d2d}]_{M\times M}$.

In this article, we focus on the uplink energy consumption of EDs, and ignore the downlink and computation energy consumption on the SCceNBs side.
On the one hand, for most computation-intensive applications, the data size of outputs sent back in the downlink is much smaller than the data size of inputs in the uplink (the ratio could be as low as 1/30 \cite{8315054}).
On the other hand, the energy consumption of uplink transmission is about 5.5 times as much as that consumed in the downlink reception.
Thus, the energy consumption of downlink reception is much less than that of uplink transmission.
In addition, the SCceNBs are usually deployed in fixed areas with sufficient energy supply.

\subsubsection{Service Caching Model}
As stated earlier, SCceNBs and EDs have storage space to store data associated with specific computing services.
Service is an abstraction of mobile applications that is managed by servers (i.e., remote cloud, SCceNBs and EDs in this work) and requested by EDs.
Running a particular service (e.g., mobile gaming, virtual reality and Google translator) requires caching the related data (e.g., libraries and databases) on the servers.
Let $\mathcal{Q}=\{1,2,...,Q\}$ represents a library which consists of $Q$ services.
Note that the services are cached at one or multiple servers (i.e., remote cloud, SCceNBs and EDs).
Each service $q$ ($q\in\mathcal{Q}$) requires a storage space $c_{q}$.
Note that the SCceNBs and EDs have limited storage capacities storing part of the popular services, whereas the remote cloud server has a infinite-capacity storage storing all the $Q$ services.
Let weight matrices $\boldsymbol{Q}_{SC}=[k_{q}^{s}]_{1\times Q}$ $(q=1,2,...,Q)$ represent the service caching placement index for the SCceNBs (i.e., all the SCceNBs share the same service caching placement scheme $\boldsymbol{Q}_{SC}$).
Each element $k_{q}^{s}\in \{0,1\}$ denotes the service caching placement index for the service $q$, $k_{q}^{s}=1$ means the service $q$ is cached in the SCceNBs, $k_{q}^{s}=0$ means the service $k$ is not cached.
Similarly, let weight matrices $\boldsymbol{Q}_{ED}=[k_{m, q}^{u}]_{M\times Q}$ $(m=1,...,M; q=1,2,...,Q)$ represent the service caching placement index for the EDs.
Each element $k_{m, q}^{u}\in \{0,1\}$ denotes the service caching placement index for the ED $m$, $k_{m, q}^{u}=1$ means the service $q$ is cached in the ED $m$, $k_{m, q}^{u}=0$ means the service $k$ is not cached.
In this work, we consider a more practical service caching placement scenario.
Thus, the matrix $\boldsymbol{Q}_{SC}$ is centralized and dynamically maintained by the I-UDEC controller (i.e., the macro base station), and the matrix $\boldsymbol{Q}_{ED}$ is distributed and statically managed by the EDs.

\subsubsection{Task Model}

\begin{figure}[t!]
    \centering
    \includegraphics[width=3in]{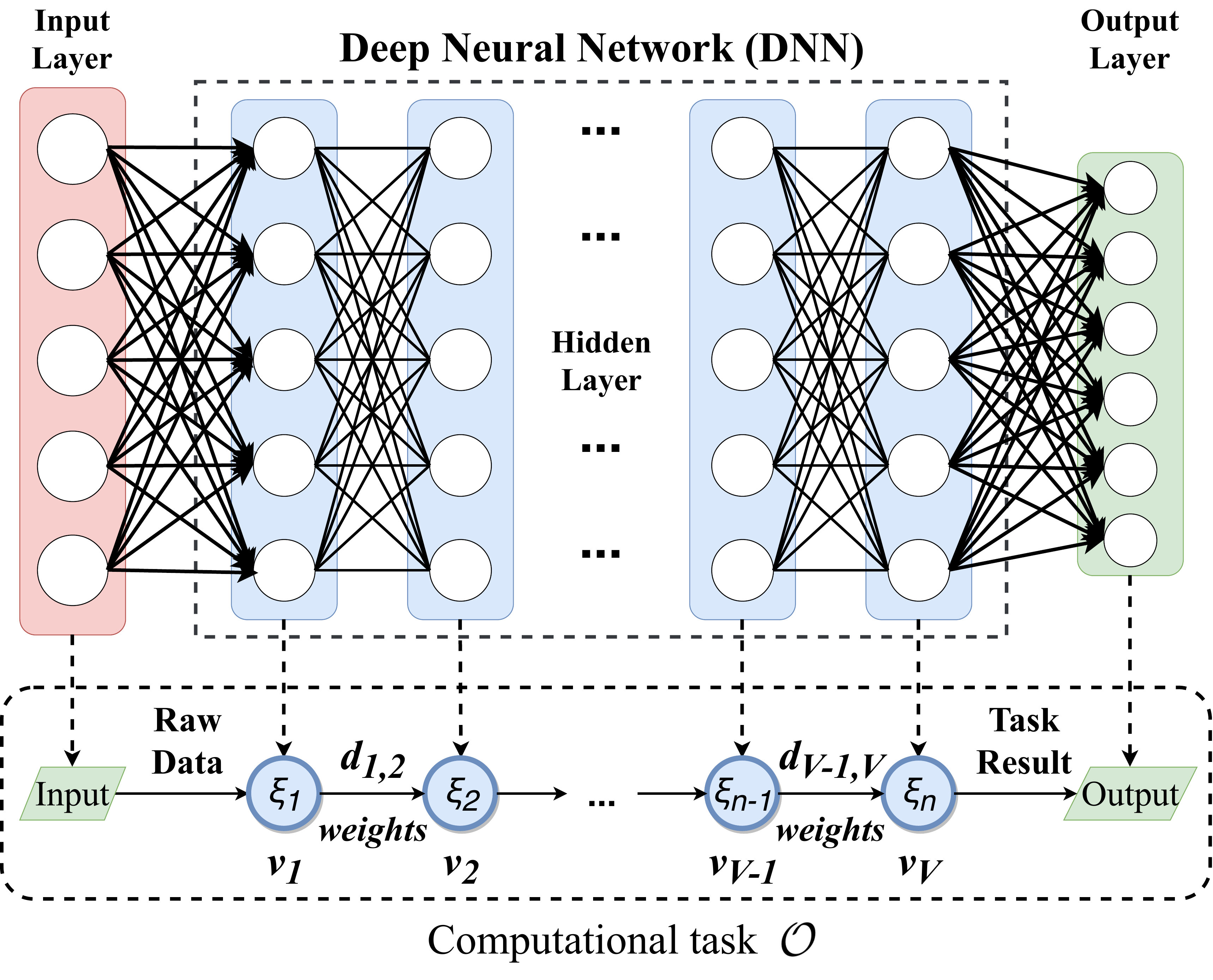}
    \caption{An illustrative task model of DNN model partition~\cite{8736011}.}
    \label{fig:DNN}
\end{figure}

In this article, we consider a fine-grained task partitioning manner~\cite{MAUI}, and split a computational task $\mathcal{O}$ into multiple sub-tasks.
Let $\mathcal{O}=(\mathcal{V},\mathcal{D})$, where $\mathcal{V}$ refers to the set of sub-tasks, $\mathcal{D}$ denotes the data dependencies between the sub-tasks.
Assume that $V$ represents the number of sub-tasks in task $\mathcal{O}$ (i.e., $V=|\mathcal{V}|$).
Each sub-task $v$ $(v\in \mathcal{V})$ can be: i) offloaded to a cloud server, i) offloaded to an edge server, or iii) executed locally in the edge device.
Note that the outputs of some sub-tasks are the inputs of others, thus, the dependencies has great influence on the task execution and computation offloading, and can not be ignored.
Generally, there exists three typical dependency models: i) sequential, ii) parallel and iii) general dependency~\cite{7463066}.
Take the task of deep neural network (DNN) model partition~\cite{8736011} with sequential dependency as an example.
In order to reduce the burden of edge intelligence applications execution on end devices, DNN model partition can split a DNN training phase (i.e., the computational task $\mathcal{O}$) into multiple sub-tasks (i.e., the hidden layers of DNN), as shown in Fig.~\ref{fig:DNN}.
By dynamically offloading the resource intensive sub-tasks to the edge server, DNN model partition can speed up the inference process.
In this work, we consider the sequential dependency task model in our framework.

Let a parameter tuple $k_{v}=\left \langle v, \xi_{v}, d_{u,v}, d_{v, w}, q(v)\right \rangle$ characterize the subtask $v$ of task $\mathcal{O}$, where $v$ is the current sub-task, $u$ and $w$ refer to the previous sub-task and next sub-task of $v$, $\xi_{v}$ ($v\in\mathcal{V}$) denotes the workload of sub-task $v$, $q(v)$ represent the required service of sub-task $v$.
$d_{u,v}$ and $d_{v, w}$ refer to the input and output data size of sub-task $v$ (e.g., weights of the DNN model), respectively.
Let $\rho_{v}$ (in CPU cycles/byte) denote the required cpu cycles a CPU core will perform per byte for the input data processed by the sub-task $v$ (i.e., complexity of sub-task $v$).
Thus, we have $\xi_{v}=\rho_{v} \cdot d_{u,v}$.

\subsection{Computation Offloading in I-UDEC}
In this article, the process of computation offloading in the I-UDEC framework consists of the following phases: i) service caching placement, ii) computation offloading decisions and iii) resource allocation.
Firstly, due to the limited storage capabilities of SCcsNBs, an optimal service caching placement strategy for the the SCcsNBs is required to shorten the service distance between EDs and their serving SCcsNBs.
%Note that many emerging mobile applications involve intensive computation based on data analytics, thus caching computation services and their related databases that are likely to be reused by others can further boost the computation performance of the entire MEC system~\cite{MEC_survey}.
%One typical example is mobile AR gaming~\cite{AR,ImmersiveGame}.
%Certain game rendered videos, e.g., gaming scenes, can be reused by other players.
%Thus, it makes sense to cache AR services and related databases at edge servers for providing real-time service.
However, unlike the remote cloud which equipped with huge computation and storage resource, an individual SCceNB with limited resource allows only part of the services to be cached at a time.
Thus, the key problems for service caching in I-UDEC are i) how to estimate the popularity of services and cache the popular services, ii) how to balance the tradeoff between multiple services and finite storage capacity of SCceNBs.
Secondly, a fine-grained application partitioning strategy is used to reduce task execution time and improve quality-of-service (QoE) for EDs.
Fine-grained application partitioning divides a mobile application into multiple sub-tasks, and dynamically offloads only the computation-intensive subtasks of the application.
Note that real-time and dynamic partitioning allows mobile devices to obtain the highest benefit through computation offloading, thus plays a critical role in UDEC networks.
At last, SCceNBs need to allocate computation, communication and storage resources to the offloaded subtasks.
After partitioning mobile applications, SCceNB need to allocate available system resources to process the offloaded parts.
Thus, an efficient resource allocation scheme will minimize the system resource usage, and significantly improve the experience of edge devices.
To this end, it makes sense to jointly optimize the service caching placement, application partitioning and resource allocation for the I-UDEC.

\begin{figure}[t!]
    \centering
    \includegraphics[width=3.5in]{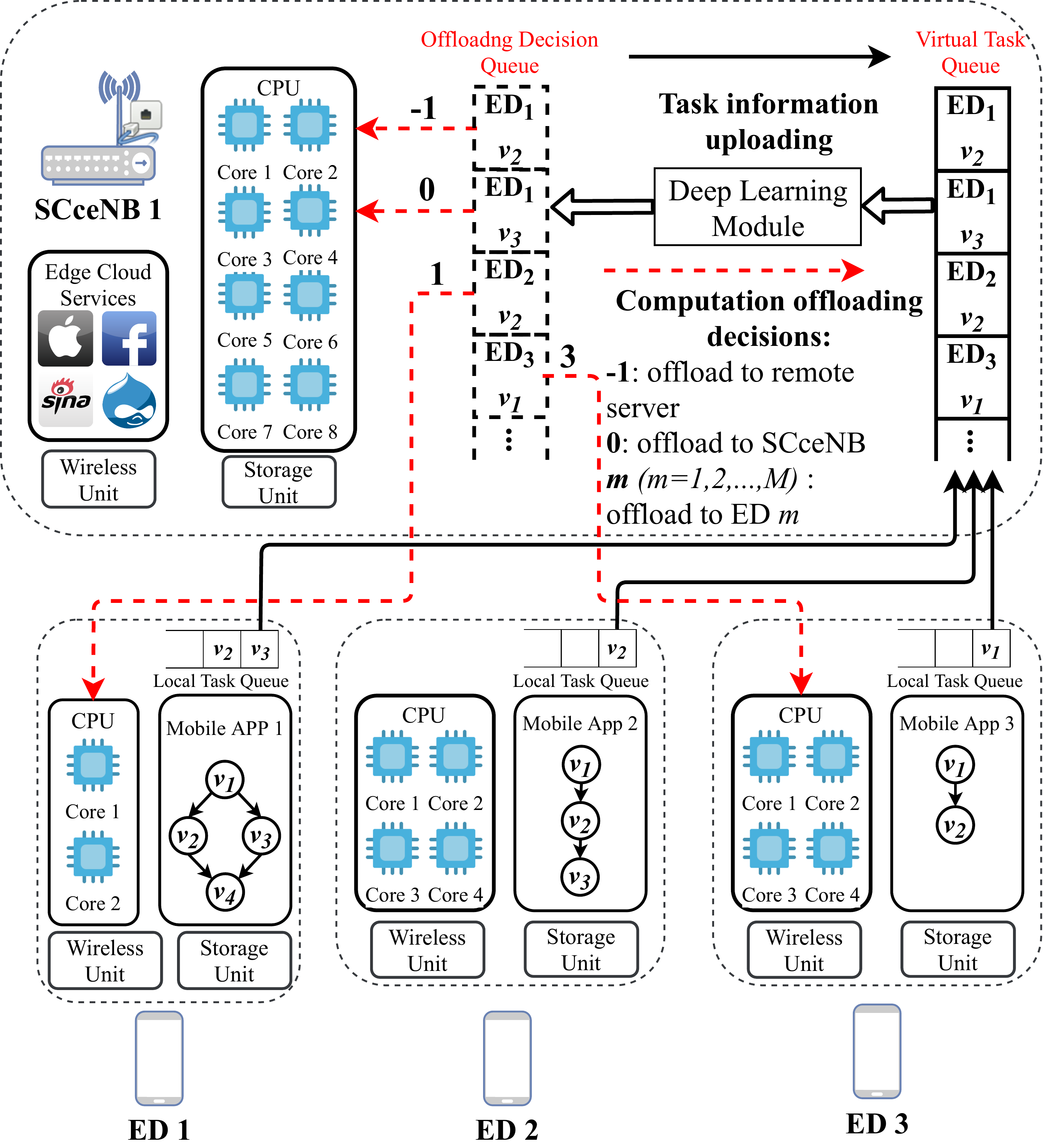}
    \caption{Computation offloading process for I-UDEC.}
    \label{fig:offloading}
\end{figure}

To this end, the computation offloading process for the I-UDEC framework is illustrated in Fig.~\ref{fig:offloading}.
In this work, we consider a hybrid computation offloading scenario for the I-UDEC, which means that an ED can either execute sub-tasks locally, or offload computation-intensive sub-tasks to i) another ED through D2D link, ii) a SCceNB through wireless uplink, or iii) remote cloud server through wireless uplink and high speed fronthaul.
Furthermore, we assume that the offloading decision making operates in a time-slot structure, and the decision making period is discrete time frames $t\in \mathcal{T}=\{0,1,2,...,T\}$.

In the I-UDEC framework, we consider each ED is served by its closest SCceNB.
Each SCceNB manages a virtual task queue to store the information of the sub-tasks pending for execution and outputs a queue of offloading decisions for edge devices, as shown in Fig.~\ref{fig:offloading}.
%The components being retrieved from the queue and processed.
Let matrix $\boldsymbol G_{n}(t)=\{\boldsymbol g_{1}(t), \boldsymbol g_{2}(t),...,\boldsymbol g_{l}(t)\}$ represents the task queue phase of SCceNB $n$ in decision period $t$, where $l$ denotes the maximum queue length for the task queue.
$\boldsymbol g_{i}(t)=\{m,k_{v}\}$ denotes a parameter vector for the $i$th element in the task queue.

Define $\boldsymbol{E}_{n}(t)=\{e_{1}(t), e_{2}(t),...,e_{l}(t)\}$ as the computation offloading action for the task queue $\boldsymbol G_{n}(t)$ in decision period $t$, where $e_{i}(t)$ $(e_{i}(t)\in \{-1, 0, 1, 2,..., l\})$ represents the offloading action for $\boldsymbol g_{i}(t)$, as shown in Fig.~\ref{fig:offloading}.
$e_{i}(t)=-1$ denotes to offload the sub-task $v$ in $\boldsymbol g_{i}(t)$ to the remote cloud server, $e_{i}(t)=0$ represents to offload the sub-task to the serving SCceNB, and $e_{i}(t)=j$ ($j=1,2,...,M$) indicates to offload the sub-task to the ED $j$.
Note that $m=j$ represents the local execution.
%Then, the offloading action index can be fed to a offloading action matrix $\boldsymbol O_{n}=[o_{i}]_{1\times l}$ for SCceNB $n$, where $l$ indicates the maximum task queue length.

Based on the above assumptions, we can obtain the total execution time of the sub-tasks in the current task queue in $\boldsymbol G_{n}(t)$ as follows:

\begin{equation}{\label{eq:exe_time}}
T^{exe}(t)=\sum_{i=1}^{l}t_{i}^{exe}(t),
\end{equation}
where $ t_{i}^{exe}$ represents the execution time for the $i$th element $\boldsymbol g_{i}(t)=\{m,k_{v}\}$ in the current task queue $\boldsymbol G_{n}(t)$ as follows:
\begin{equation}{\label{eq:exe_time_sub}}
t_{i}^{exe}(t)=
\begin{cases}
\frac{\rho_{v} \cdot d_{u,v}}{X_{m}\cdot f_{u}}, &e_{i}(t)=j, 1\leq j\leq M, j=m;\\
\\
\frac{\rho_{v} \cdot d_{u,v}}{X_{j}\cdot f_{u}}+\frac{d_{u,v}}{r_{m,j}^{d2d}}, &e_{i}(t)=j, 1\leq j\leq M, j\neq m, \\
&k_{j, q(v)}^{u}=1;\\
\\
\frac{\rho_{v} \cdot d_{u,v}}{f_{s}}+\frac{d_{u,v}}{r_{m,n}^{ul}}, &e_{i}(t)=0, k_{q(v)}^{s}=1;\\
\\
\frac{d_{u,v}}{r_{m,n}^{ul}}+t_{n}^{e}, &e_{i}(t)=-1, \text{else}.\\
\end{cases}
\end{equation}
Note that $k_{j, q(v)}^{u}=1$ guarantees the requires service $q(v)$ of subtask $v$ is cached in ED $j$, if ED $m$ decides to offload his subtask subtask $v$ to ED $j$ through D2D offloading.
Similarly, $k_{q(v)}^{s}=1$ guarantees requires service $q(v)$ of subtask $v$ is cached in SCceNB $n$, if ED $m$ performs edge offloading (i.e., offload his subtask to the nearby SCceNB $n$).
$t_{n}^{e}$ refers to the end-to-end latency between SCceNB $n$ and the remote cloud server, when ED $m$ performs cloud offloading.

\section{Problem Formulation and Algorithm Design}
\begin{figure}[t!]
    \centering
    \includegraphics[width=3.5in]{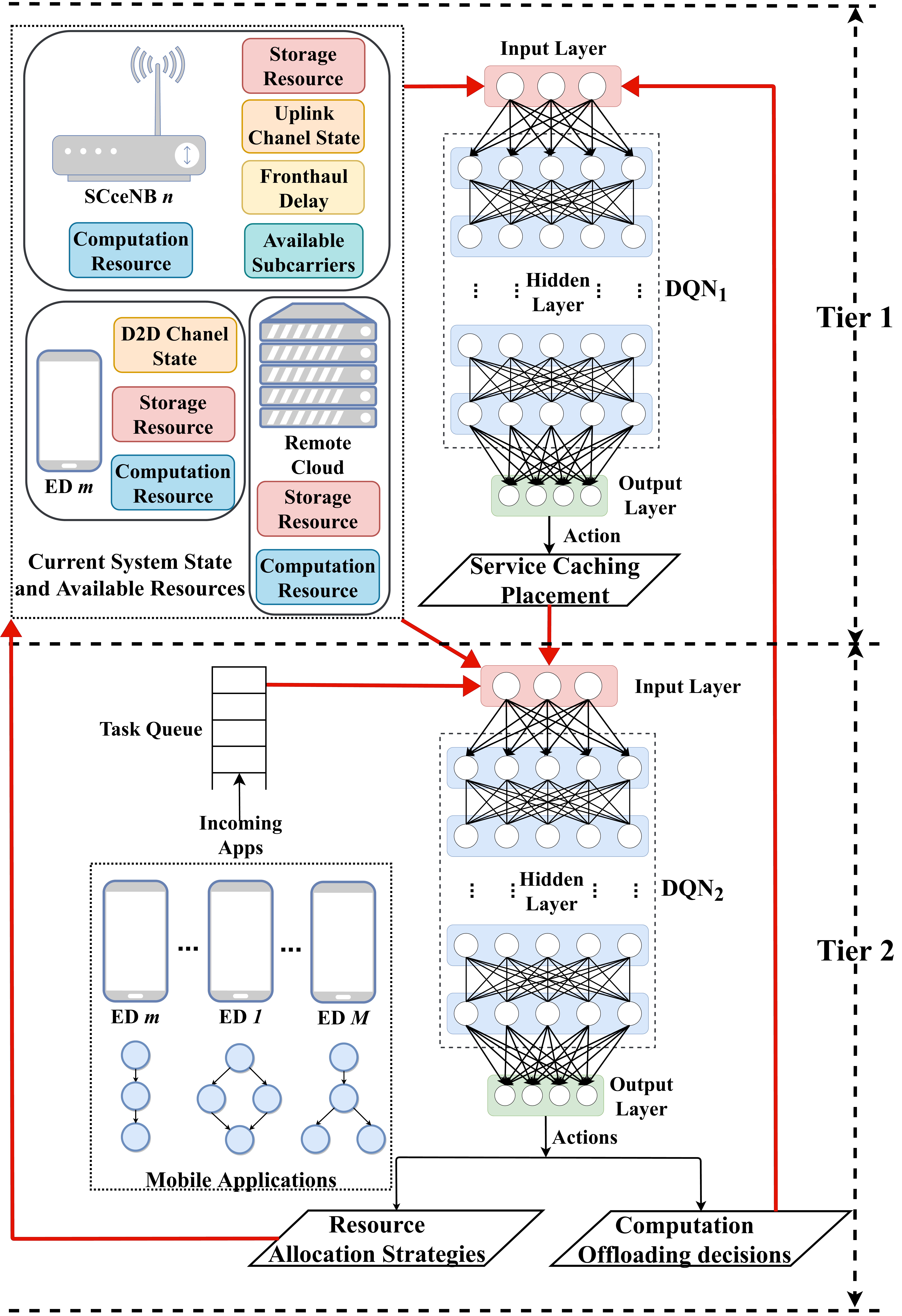}
    \caption{Proposed two-timescale deep reinforcement learning (\textit{2Ts-DRL}) framework.}
    \label{fig:framework}
\end{figure}

{\label{sec:Problem_Formulation}}
In this section, we will formulate the optimization problem of joint application partitioning, resource allocation and service caching placement in the I-UDEC framework.
The main objective of the optimization problem is to jointly minimize the tasks' long-term execution time and system resources' usage (i.e., computation, communication and storage usage).
To this end, we utilize deep Q-learning (DQL) algorithm to tackle the formulated optimization problem, and present a novel two-timescale deep reinforcement learning (\textit{2Ts-DRL}) approach to train the decision agent, as shown in Fig.~\ref{fig:framework}.

In reinforcement learning (RL)~\cite{9062302}, an agent can periodically learn to take actions, observes the most reward and automatically adjusts its strategy in order to obtain the optimal policy.
For example, Q-learning~\cite{8976180} is a widely used model-free RL algorithm in the literature.
In Q-learning, decision agent can reach to the best policy by estimating the Q-function (be defined as the state-action function).
Q-function utilizes the samples which are obtained during the interactions with the environment to approximate the values of state-action pairs.
However, it is time-consuming for RL to obtain the best policy, since RL must explore and gain knowledge of the whole environment.
Thus, RL is unsuitable and inapplicable to large-scale networks, especially for the ultra dense network environment in the I-UDEC.
Deep learning is introduced as a state-of-the-art technique to tackle the above limitation of RL, and opens a novel research field, namely deep reinforcement learning (DRL).
DRL improves the learning speed and reinforcement learning performance by leveraging deep neural networks (DNNs) to train the learning model.

Deep Q-Learning (DQL) algorithm which integrates deep learning into reinforcement learning, is introduced to solve the problem of small state space and action space in Q-learning.
The DQL implements a Deep Q-Network (DQN) instead of the Q-table of Q-learning, thus, the learning speed is significantly improved.

Note that, in the I-UDEC framework, the application partitioning, resource and service caching placement strategies have different delay sensitivities.
To this end, we present a novel two-timescale deep reinforcement learning (DRL) method to jointly optimize the above issues in two different timescale.
Specifically, we design a two-tier structure deep-Q network (DQN) agent for the DRL, which operates in two different timescales, as shown in Fig.~\ref{fig:framework}.
The bottom tier of the agent outputs delay sensitive decisions (i.e., application partitioning and resource allocation strategies) in a fast timescale, whereas the top tier outputs delay insensitive decision (i.e., service caching placement strategy) in a slow timescale.

Specifically, we  model the problem as a Markov decision process (MDP) \cite{MDP2}, and formulate the state space, action space and reward function of the MDP as follows.
\subsection{State Space}
At the beginning of each decision period $t$, the EDs send the information of their computation offloading requests, current service caching placement strategy and available resource to their nearby SCceNBs and the controller (i.e., the macro base station in Fig.~\ref{fig:I-UDEC}).
Then, each SCceNB receives the uploaded information, and builds a virtual task queue $\boldsymbol G_{n}(t)$ ($n\in \mathcal{N}$).
Thus, the system state of SCceNB $n$ at each decision period $t$ is composed of the task queue state and the available resource state, which is given as follows:
\begin{equation}{\label{eq:state}}
\begin{split}
\boldsymbol S_{n}(t)=&\{{K}(t),\boldsymbol R_{n}^{UL},\boldsymbol R_{n}^{D2D},\boldsymbol Q_{SC}(t),\boldsymbol Q_{ED}(t),\boldsymbol G_{n}(t),\boldsymbol X_{n}(t),\\
&\boldsymbol Y_{n}(t),\boldsymbol Z_{n}(t)\},
\end{split}
\end{equation}
%where $\boldsymbol C^{s}(t)=[c_{1}^{s}(t), c_{2}^{s}(t),..., c_{Y}^{s}(t)]$ represents the state of available computation resource (i.e., CPU cores) for the SCceNB $n$.
%Each element $c_{i}^{s}(t)\in \{0,1\}$ $(i=1,2,...,Y)$ denotes the state for the $i$th CPU core of the SCceNB, $c_{i}^{s}(t)=1$ means the CPU core $i$ is available at decision period $t$, $c_{i}^{s}(t)=0$ means the CPU core is not available.
%$\boldsymbol C^{u}(t)=[c_{1}^{u}(t), c_{2}^{u}(t),..., c_{M}^{u}(t)]$ represents the state of available computation resources for the $m$ EDs.
%Similarly, each element $c_{j}^{u}(t)\in \{0,1\}$ $(j=1,2,...,M)$ denotes the state of computation resource for the $j$th ED, $c_{j}^{u}(t)=1$ means the computation resource of ED $j$ is available in the decision period $t$, $c_{j}^{u}(t)=0$ means the computation resource of ED $j$ is not available.
where $K(t)$ denotes the available subcarriers to be allocated at decision period $t$.
$\boldsymbol R_{n}^{UL}$ and $\boldsymbol R_{n}^{D2D}$ are the uplink and d2d data rate weight matrices, respectively.
$\boldsymbol Q_{SC}(t)$ and $\boldsymbol Q_{ED}(t)$ denote the service caching placement state for the SCceNB $n$ and $M$ EDs at decision period $t$, respectively.
$\boldsymbol G_{n}(t)$ refers to the current task queue state.
$\boldsymbol Y_{n}(t)=\{y_{1}(t),y_{2}(t),...,y_{Y}(t)\}$ is the current computation resource (i.e., CPUs) state, where $y_{i}(t)=1$ $(i=1,2,...,Y)$ represents the $i$th CPU core of SCceNB $n$ is available, and $y_{i}(t)=0$, not available.
Similarly, $\boldsymbol X_{n}(t)=\{x_{1}(t),x_{2}(t),...,x_{M}(t)\}$ is the current computation resource the $M$ EDs within the coverage of SCceNB $n$, where $x_{j}(t)=1$ $(j=1,2,...,M)$ represents the computation resource of the $j$th ED is available, and $x_{j}(t)=0$, not available.
$\boldsymbol Z_{n}(t)=\{q_{v}, v\in \boldsymbol G_{n}(t)\}$ represent the set of required services for the subtasks in the task queue.
\subsection{Action Space}
In the I-UDEC framework, the decision agent has to take the following three actions: i) application partitioning actions $\boldsymbol{E}_{n}(t)$ for the current task queue $\boldsymbol G_{n}(t)$, ii) wireless resource (i.e., subcarrier) allocation strategy $\boldsymbol K_{n}(t)=\{K_{1}^{n}(t),K_{2}^{n}(t),...,K_{M}^{n}(t)\}$, where $K_{m}^{n}(t)$ $(m=1,2,...,M)$ represents the number of subcarriers that allocates to the ED $m$ at decision period $t$,
and iii) service caching placement policy $\boldsymbol Q_{SC}(t)$ for the SCceNBs.
Note that $\boldsymbol E_{n}(t)$ and $\boldsymbol K_{n}(t)$ are delay intensive actions, which must be made immediately (i.e., real-time resource scheduling).
Whereas $\boldsymbol Q_{SC}(t)$ is a delay insensitive action, because SCceNBs do not have to frequently update their cached services.
Thus, we assume a time interval $\alpha t$ for SCceNB $n$ to update their cached services.
It means that the SCceNBs update their service caching strategy $\boldsymbol Q_{SC}(t)$ every $\alpha t$ seconds.
%Based on the observed composite state $\boldsymbol S_{n}(t)$, SCceNB $n$ allocates computation resources (i.e., computation offloading) and communication resources (i.e., subcarriers) to the sub-tasks in current task queue.
%Note that, due to system resource scheduling delay, SCceNB $n$ must rent the resources for a long enough time $\Delta t$, to ensure that the EDs can finish executing any application.

\subsection{Reward Function}
In order to minimize the overall task execution time for the EDs and reduce the system resource usage of I-UDEC, we consider a comprehensive revenue of the I-UDEC framework as the system's reward.
The revenue combines of the computation, communication and storage resources usage, as well as the task execution time for the EDs.

For the delay insensitive service caching placement strategy (i.e., tier 1 in Fig.~\ref{fig:framework}), given the system state $\boldsymbol S_{n}(t)$, service request history $\boldsymbol{Z}_{n}(t-\alpha t,t-1)$, current service caching placement strategies $\boldsymbol Q_{SC}(t)$, $\boldsymbol Q_{ED}(t)$ and resource allocation strategy $\boldsymbol K_{n}(t)$ at current decision period $t$, the immediate reward $R_{s}(t)$ of service caching can be expressed as follows:
\begin{equation}
\begin{aligned}
\label{eq:reward_cache}
R_{c}(t)=&\zeta_{1}\frac{T^{exe}(t)}{T^{loc}(t)}+\zeta_{2}\frac{\sum_{q=1}^{Q}k_{q}^{s}}{C_{s}},\\
    \textit{s.t.\quad}
    &T^{exe}(t)\leqslant T^{loc}(t),\\
    &\sum_{q=1}^{Q}k_{q}^{s} \leqslant C_{s},\\
\end{aligned}
\end{equation}
where $\zeta_{1}$ and $\zeta_{2}$ denote weight factors.
$T^{loc}(t)$ is the total local execution time for the subtask in the current task queue as follows:
\begin{equation}{\label{eq:T_local}}
T^{loc}(t)=\sum_{i=1}^{l}\frac{\rho_{v} \cdot d_{u,v}}{X_{m}\cdot f_{u}}, v\in \boldsymbol g_{i}(t).
\end{equation}
Note that we use the service request history $\boldsymbol{Z}_{n}(t-\alpha t,t-1)$ to estimate the popularity of the services, and the SCceNBs prone to cache the most popular services to enhance their cache hit probability.
The first term of $R_{c}(t)$ in (\ref{eq:reward_cache}) is a normalized execution time index, and the first constraint of (\ref{eq:reward_cache}) guarantees the total execution time through computation offloading must less than the time of local processing.
Similarly, the term second of $R_{c}(t)$ in (\ref{eq:reward_cache}) represents a normalized storage usage index, and the second constraint of (\ref{eq:reward_cache}) guarantees the total storage consumption of current service caching strategy $\boldsymbol Q_{SC}(t)$ must less than the storage capacities of the SCceNBs.
The objective of the service caching placement is to minimize the long-term execution time for the EDs as well as the storage usage for the SCceNBs.
Since the objective of DRL is to maximize reward, the value of long-term reward should be negatively correlated to the immediate reward as follows:
\begin{equation}{\label{eq:reward_long}}
R_{c}^{long}=max \sum -R_{c}(t),
t=0,\alpha t,2\alpha t,...,T.
\end{equation}

For the application partitioning and resource allocation strategies (i.e., tier 2 in Fig.~\ref{fig:framework}), given the system state $\boldsymbol S_{n}(t)$, current service caching placement strategies $\boldsymbol Q_{SC}(t)$, $\boldsymbol Q_{ED}(t)$, available communication resource $K(t)$, available computation resource $\boldsymbol Y_{n}(t)$ and $\boldsymbol X_{n}(t)$ at current decision period $t$, the immediate reward $R_{a}(t)$ of application partitioning and resource allocation can be expressed as follows:
\begin{equation}{\label{eq:reward_ar}}
R_{a}(t)=\mu_{1}\frac{T^{exe}(t)}{T^{loc}(t)}+\mu_{2}\frac{\sum_{m=1}^{M}K_{m}^{n}(t)}{K(t)}+\mu_{2}\frac{\sum_{j=1}^{l}e_{j}(t)=0}{\sum_{i=1}^{Y}y_{i}(t)},
\end{equation}
where $\mu_{1}$, $\mu_{2}$ and $\mu_{3}$ denote weight factors.
The first term in (\ref{eq:reward_ar}) represents a normalized execution time index, as shown in (\ref{eq:reward_cache}).
The second term in (\ref{eq:reward_ar}) denotes a normalized communication resource usage of SCceNB $n$, which guarantees the number of consumed subcarriers must less than the number of available subcarriers at decision period $t$.
The third term in (\ref{eq:reward_ar}) is a normalized computation resource usage of SCceNB $n$, which guarantees the number of occupied CPUs must less than the number of available CPUs of SCceNB $n$ at decision period $t$.
By leveraging DQN, we can obtain an optimal policy to maximize the long-term reward, the cumulative reward should be negatively written as follows:
\begin{equation}{\label{eq:reward_long1}}
R_{a}^{long}=max \sum -R_{a}(t),
t=0,1,2,...,T.
\end{equation}
Note that (\ref{eq:reward_long}) and (\ref{eq:reward_long1}) have a similar composition, but they operate in different timescale.

At last, our decision agent will output near-optimal resource allocation strategies, computation offloading decisions and service caching placement strategies to minimize the long-term cost.
Service caching allows for a long updating interval while computation offloading needs fast response.

\begin{figure}[t!]
    \centering
    \includegraphics[width=2.5in]{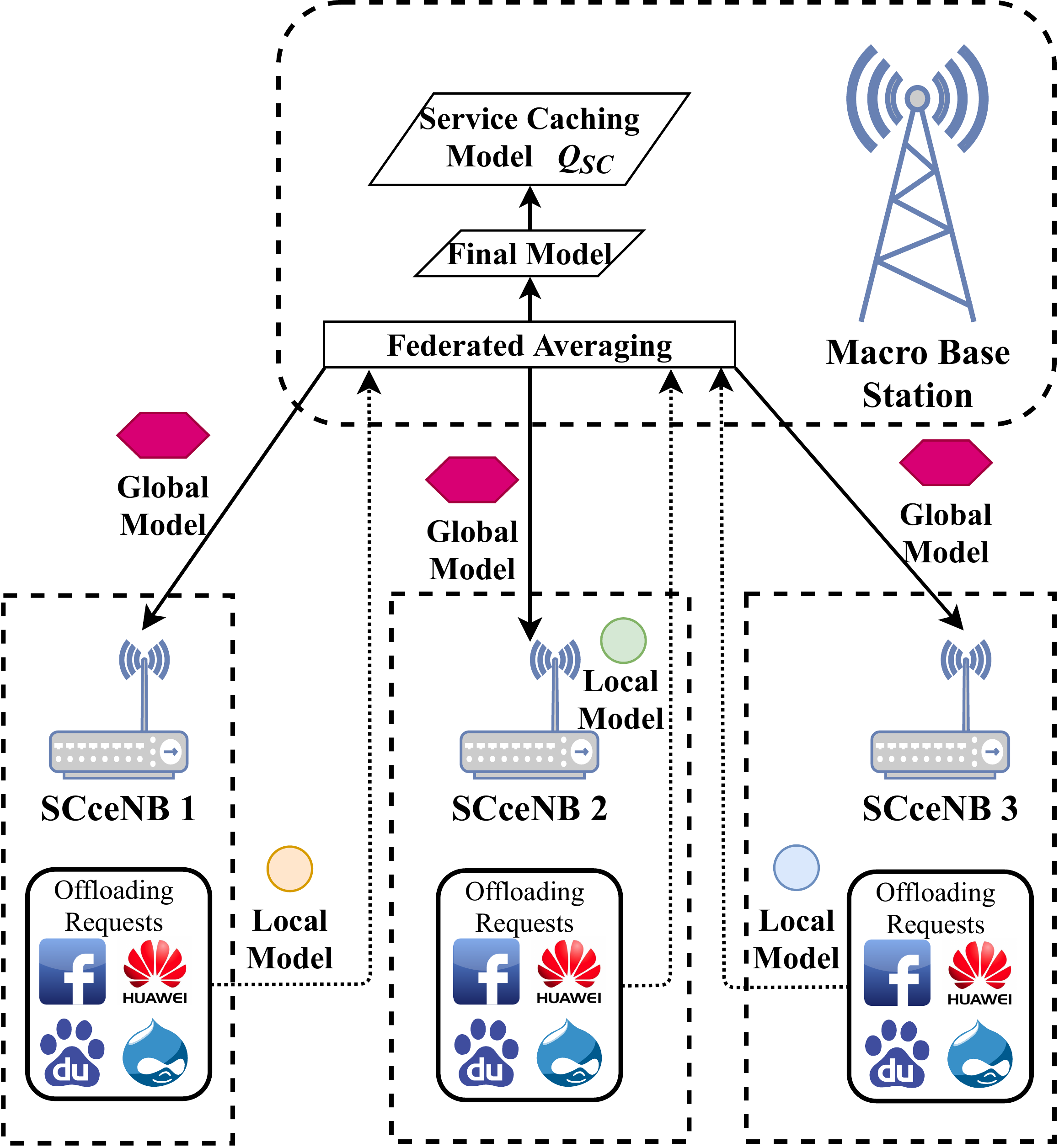}
    \caption{Federated learning (FL) based model training.}
    \label{fig:FL}
\end{figure}

\section{Federated Learning-based Model Training}{\label{sec:Federated_Learning}}
For the service caching placement agent of the proposed two-tier DQN (i.e., tier 1 in Fig.~\ref{fig:framework}), it is critical to predict the future popularity of services.
Thus, the most popular ones can be cached to minimized the long-term reward in (\ref{eq:reward_long}).
Thus, the controller (i.e., the macro base station in Fig.~\ref{fig:I-UDEC}) collects the service request history $\boldsymbol{Z}_{n}(t-\alpha t,t-1)$ to make the prediction.
However, it may bring a significant challenge: EDs may not trust the controller and thus unwilling to transmit the information of their offloading requests (i.e., most used services).

To this end, we design a federated learning-based model training method in order to distributively train the service caching placement agent, as shown in Fig.~\ref{fig:FL}.
Algorithm~\ref{alg:FL} shows the pseudocode of the proposed algorithm.
Specifically, at the beginning of decision period $t$, all the $N$ SCceNBs firstly receive the global model $\boldsymbol W(t)$ (i.e., DRL weights) according to their current service placement strategy $\boldsymbol Q_{SC}(t)$ from the central controller.
Then, the SCceNBs compute the local models $\boldsymbol W_{1}(t), \boldsymbol W_{2}(t),...,\boldsymbol W_{N}(t)$ based on their current service request $\boldsymbol Z_{n}(t), (n=1,2,...,N)$.
Next, each SCceNB $n$ uploads their updated model (be written as $\boldsymbol H_{n}(t):=\boldsymbol W(t)-\boldsymbol W_{n}(t))$ to the central controller.
The controller receives and aggregates the updates, and then constructs an updated global model $\boldsymbol W(t+1)$ in the next decision period by using federated averaging~\cite{DBLP:conf/nips/SmithCST17,9062302} as follows:

\begin{equation}{\label{eq:federated_averaging}}
\boldsymbol W(t+1)=\boldsymbol W(t)+\psi \boldsymbol H(t),
\boldsymbol H(t):=\frac{1}{N}\sum_{n=1}^{N}\boldsymbol H_{n}(t),
\end{equation}
where $\psi$ represents the learning rate.

\begin{algorithm}[t]
\caption{The federated learning-based model training for service caching placement.}
\label{alg:FL}
  \begin{algorithmic}[1]
    \Require\\
    Controller side:

    Initialize the DRL model with random weights $\boldsymbol W(0)$ at the beginning of decision period $t=0$.
    \\
    SCceNBs' side:

    Initialize the local DRL model weights $\boldsymbol W_{n}(0)$, $(n=1,2,...,N)$;

    Download $\boldsymbol W(0)$ from the controller and let $\boldsymbol W_{n}(0)=\boldsymbol W(0)$, $(n=1,2,...,N)$.
    \Ensure
      for each decision period $t=0$ to $T$ do
    \Function{FL}{$\boldsymbol{Z}_{n}(t)$}

    SCceNBs' side:

    \While{$t>0$}
        \For{each SCceNB $n\in\mathcal{N}$ in parallel}
             \State download $\boldsymbol W(t)$ from the controller;
             \State let $\boldsymbol W_{n}(t)=\boldsymbol W(t)$;
             \State train the DRL agent locally with $\boldsymbol W_{n}(t)$ on the current service requests $\boldsymbol{Z}_{n}(t)$;
             \State Upload the trained weights $\boldsymbol W_{n}(t+1)$ to the controller;
        \EndFor

     Controller side:
    \State receive all weights $\boldsymbol W_{n}(t)$ updates;
    \State perform federated averaging;
    \State broadcast averaged weights $\boldsymbol W_{n}(t+1)$;
    \EndWhile
    \EndFunction
  \end{algorithmic}
\end{algorithm}

The centralized controller aggregates all updated models by utilizing the weighted average sum.
At the same time, the controller considers the quantity of each SCceNB's local dataset.
The iteration repeats until it reaches to the next service caching placement decision period $t+\alpha t-1$.
The controller generates a service caching placement strategy $\boldsymbol Q_{SC}(t+\alpha t)$ for the next caching placement decision period $t+\alpha t$ and broadcasts the strategy to the SCceNBs.
Then, the SCceNBs update their local cached services according to the received caching placement strategy and fetch the services from backbone network.
At last, the SCceNBs enter the next decision period and train the fast timescale agents (as shown in Fig.~\ref{fig:framework}) asynchronously in parallel upon current system state.

In addition, we use Asynchronous Advantage Actor-Critic (A3C)~\cite{9014268} to train the agents, owing to its asynchronous advantage.
Under the field of DRL, A3C uses asynchronous multi-threads to train DNN through multiple agents.
The agents are controlled by a global network, can learn and run asynchronously, thus keep their own network parameters and the copies of the environment with different policies.
Each time the agents first update their own parameters of the actor network and the critic network (i.e., the local model in Fig.~\ref{fig:FL}), and also a global actor network and a global critic network (i.e., the global model in Fig.~\ref{fig:FL}).
Then, they submit the parameters to the global networks.
At last, the global network receives and distributes the received parameters.
The local and global networks will converge after multiple iterations.
This kind of training has the following advantages:
i) lower memory usage: since the replay memory (e.g., Q-table) of standard DRL is not needed,
ii) avoiding the correlation: since different agents will likely experience different states and transitions
and iii) faster and more robust than the traditional DRL approach.

\section{performance evaluation}{\label{sec:performance_evaluation}}
In this section, the performance of the proposed \textit{2Ts-DRL} is evaluated through computer simulations.
Specifically, we use MATLAB reinforcement learning toolbox and deep learning toolbox~\cite{matlab_rl} to implement the (\textit{2Ts-DRL}) framework.
First, we describe the simulation environment and introduce the related benchmark strategies for the proposed I-UDEC.
Then, we compare the performance of the proposed \textit{2Ts-DRL} algorithm with the benchmark strategies and discuss simulation results.
The values of the parameters are summarized in Table~\ref{tab:Parameters}.

\subsection{Simulation Environment}
In our simulation, we consider a UDEC network environment.
For the communication resource in the network, we assume that the bandwidth of each subcarrier $B$ is 15KHz, and each SCcenNB has 256 subcarriers to be allocated (i.e., $K=256$).
For the computation resource, we assume that the computation capacity $f_{u}$ (i.e., CPU frequency) of a single CPU core for the EDs is $10^{9}$ cycles/s, and the number of CPU core $X_{m}$ for ED $m$ is uniformly selected from the set $\{1,2,4,8\}$.
Similarly, let the computation capacity $f_{s}$ of a single CPU core for the SCceNBs is $10^{9}$ cycles/s, and the number of CPU core $Y$ for the SCceNBs is fixed to 8.
For the services required by the EDs, we assume that there are 30 kinds of services (i.e., $Q=30$) and the data size of each service $c_{q}$ follows the uniform distribution in the range of [100, 300] M.
We also consider a popularity-based caching placement policy~\cite{Zipf}, thus the request probability for a service $q$ ($q\in\mathcal{Q}$) is:

\begin{equation}{\label{eq:Zipf}}
\begin{aligned}
f(q,\delta,Q)=\frac{1}{q^{\delta}}\sum\limits_{n=1}^{Q}\frac{1}{n^{\delta}},
\end{aligned}
\end{equation}
where $\delta$ represents the skewness of the popularity profile.

\begin{footnotesize}
\begin{table}[tp]
\caption{Network Parameters}
\label{tab:Parameters}
\centering
\renewcommand\arraystretch{0.9}
\begin{tabular}{|m{17mm}<{\centering}|m{17mm}<{\centering}||m{17mm}<{\centering}|m{17mm}<{\centering}|}\hline
Parameter&Value&Parameter&Value\\\hline
$B$              &15KHz              &$\mu_{1}$, $\mu_{2}$, $\mu_{3}$           & 1/3                  \\\hline
$M$              &5                  &$\zeta_{1}$, $\zeta_{2}$                  & 1/2                  \\\hline
$N$              &15                 &$Q$                                       & 30                   \\\hline
$f_{u}$          &$10^{8}$ cycles/s  &$c_{q}$                                   & [100, 300]M          \\\hline
$f_{s}$          &$10^{9}$ cycles/s  &$C_{s}$                                   & 2G                   \\\hline
$\rho_{v}$       &[4000, 12000] cpb  &$d_{u,v}$                                & [100, 500] KB              \\\hline
Learning rate    &0.001        &Discount factor                           & 0.95                 \\\hline
Optimizer     &Adam             &Activation function               & ReLU                \\\hline
Exploration rate     &0.1              &Hidden layer                           & 3                 \\\hline
\end{tabular}
\end{table}
\end{footnotesize}

Afterwards, we implemented our proposed \textit{2Ts-DRL} based task execution scheme and compare it with respect to the following four task execution schemes, a resource allocation strategy and a service caching placement policy, namely:

\begin{itemize}
\item \emph{Local Execution Scheme (LES):}

Local execution, all the subtasks are locally executed on the edge devices.

\item \emph{Edge Execution Scheme (EES):}

The EDs offload all their tasks to their nearby SCceNBs for edge processing.
Note that, if the required service for the subtask $v$ of ED $m$ is not cached in a nearby SCceNBs.
ED $m$ processes the subtask $v$ locally.

\item \emph{Random Execution Scheme (RES):}

Represents a random computation offloading strategy, where each subtask is randomly offload to SCceNB, cloud server, a nearby ED, or be executed locally.

\item \emph{Cloud Execution Scheme (CES):}

All the subtasks are offloaded to remote cloud server.

\item \emph{Fair Resource Allocation strategy  (FRAS):}

The communication resource (i.e., available subcarriers) are equally shared by the EDs.

\item \emph{Popularity-based Service Caching Placement Policy (PSCPP):}

The SCceNBs always cache the most popular services until reaching to their storage capacity $C_{s}$.

\end{itemize}

\subsection{Simulation Results}

\begin{figure}[t!]
    \centering
    \includegraphics[width=3in]{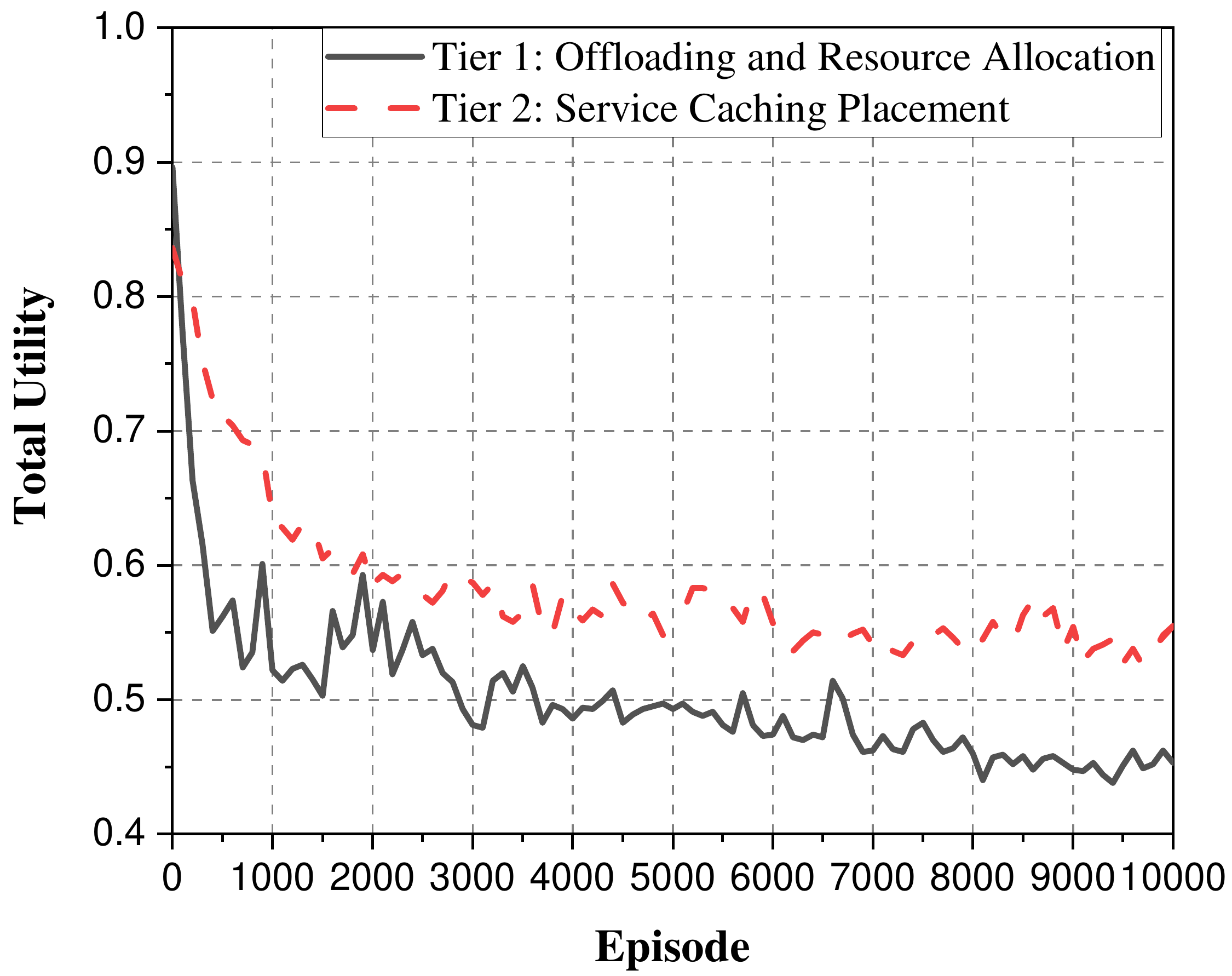}
    \caption{Convergence performance of the proposed \textit{2Ts-DRL}.}
    \label{fig:convergence}
\end{figure}

Fig.~\ref{fig:convergence} illustrates the convergence performance for the \textit{2Ts-DRL} algorithm.
Several observations can be made.
First, we can find that the total utility for both of the tiers of the \textit{2Ts-DRL} is high at the beginning of the DRL process.
The utility decreases when the number of the episodes increases.
Then, tier 1 of the \textit{2Ts-DRL} converges faster than the tier 2 of the \textit{2Ts-DRL}, since tier 1 has smaller state and action spaces.

\begin{figure}[t!]
    \centering
    \includegraphics[width=3in]{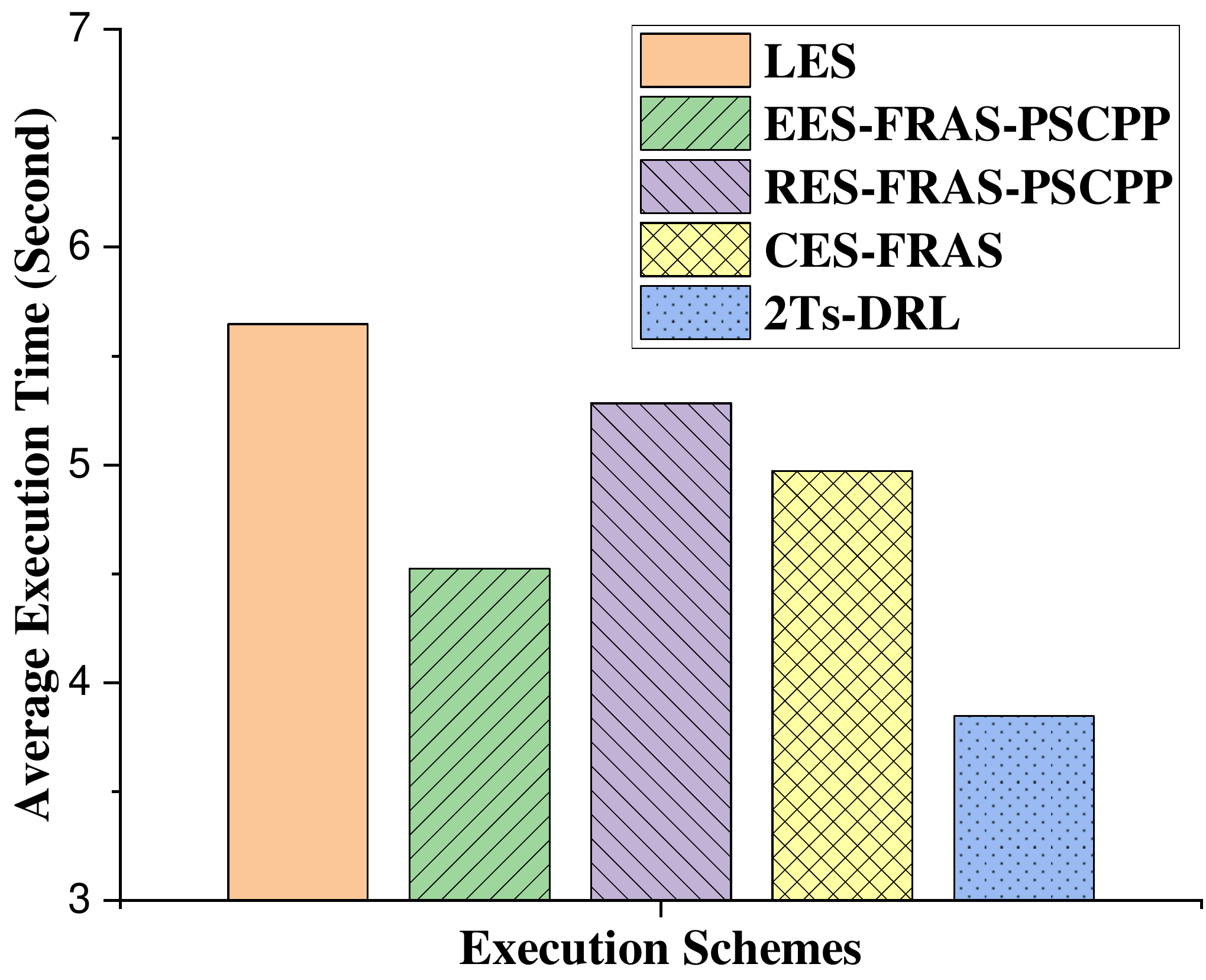}
    \caption{Task execution time for different execution schemes.}
    \label{fig:delay}
\end{figure}

Fig.~\ref{fig:delay} reports the average long-term execution time of the subtasks under different execution schemes.
Since resource allocation strategy and service caching placement policy have no impact on the LES, and service caching placement policy has have no impact on the CES.
We compare the performance of our proposed \textit{2Ts-DRL} with the benchmark strategies that is illustrated in Fig.~\ref{fig:delay}.
Specifically, EES-FRAS-PSCPP refers to an edge execution scheme with fair resource allocation strategy and popularity-based service caching placement policy.
RES-FRAS-PSCPP denotes a random execution scheme with fair resource allocation strategy and popularity-based service caching placement policy.
CES-FRAS represents a cloud execution scheme with fair resource allocation strategy when EDs upload tasks to their nearby SCceNBs.
Note that the \textit{2Ts-DRL} outperforms other execution schemes in execution time saving.
In fact, our \textit{2Ts-DRL} can reduce the task execution time on average by $31.87\%$, $14.96\%$, $27.16\%$ and $22.61\%$ compared to the LES, EES-FRAS-PSCPP, RES-FRAS-PSCPP and CES-FRAS schemes, respectively.

\begin{figure}[t!]
    \centering
    \includegraphics[width=3in]{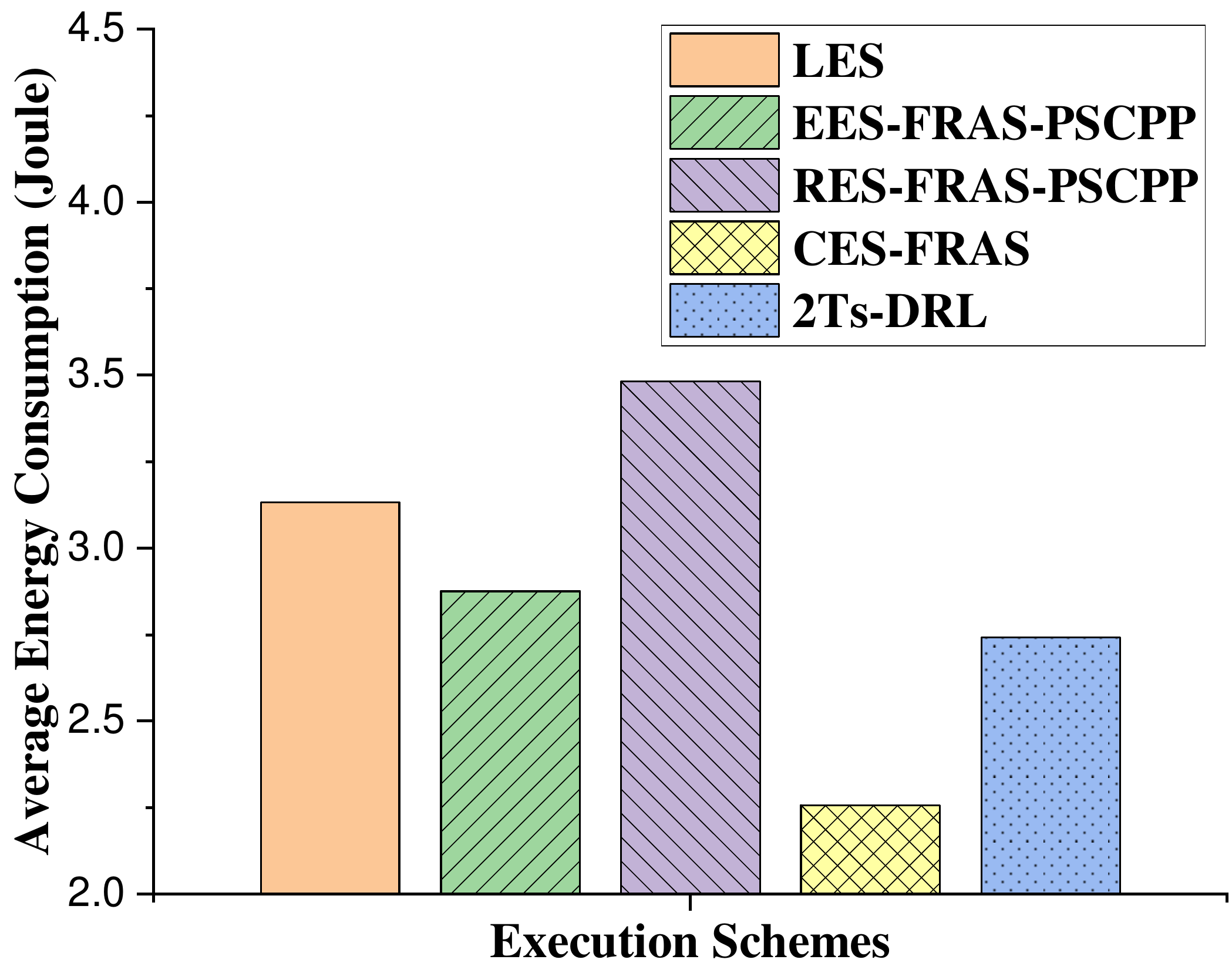}
    \caption{Energy consumption for different offloading schemes when wireless link is intermittent.}
    \label{fig:energy}
\end{figure}

Fig.~\ref{fig:energy} illustrates the average energy consumption of the subtask execution.
Note that, we only consider the energy consumption on the edge devices and ignore the execution energy consumption on the SCceNBs and remote cloud server.
Thus, CES-FRAS consumes the least amount of energy, since all the sub-tasks are offload to the remote cloud server and EDs only consume the energy for data transmission.
However, due to the long end-to-end delay, the execution time is long with respect to the \textit{2Ts-DRL}, as shown in Fig.~\ref{fig:delay}.
Note that our \textit{2Ts-DRL} outperforms EES-FRAS-PSCPP and RES-FRAS-PSCPP, and achieves $25.15\%$, $17.25\%$ and $10.57\%$ energy reduction compared to the above execution schemes.

\begin{figure}[t!]
    \centering
    \includegraphics[width=3in]{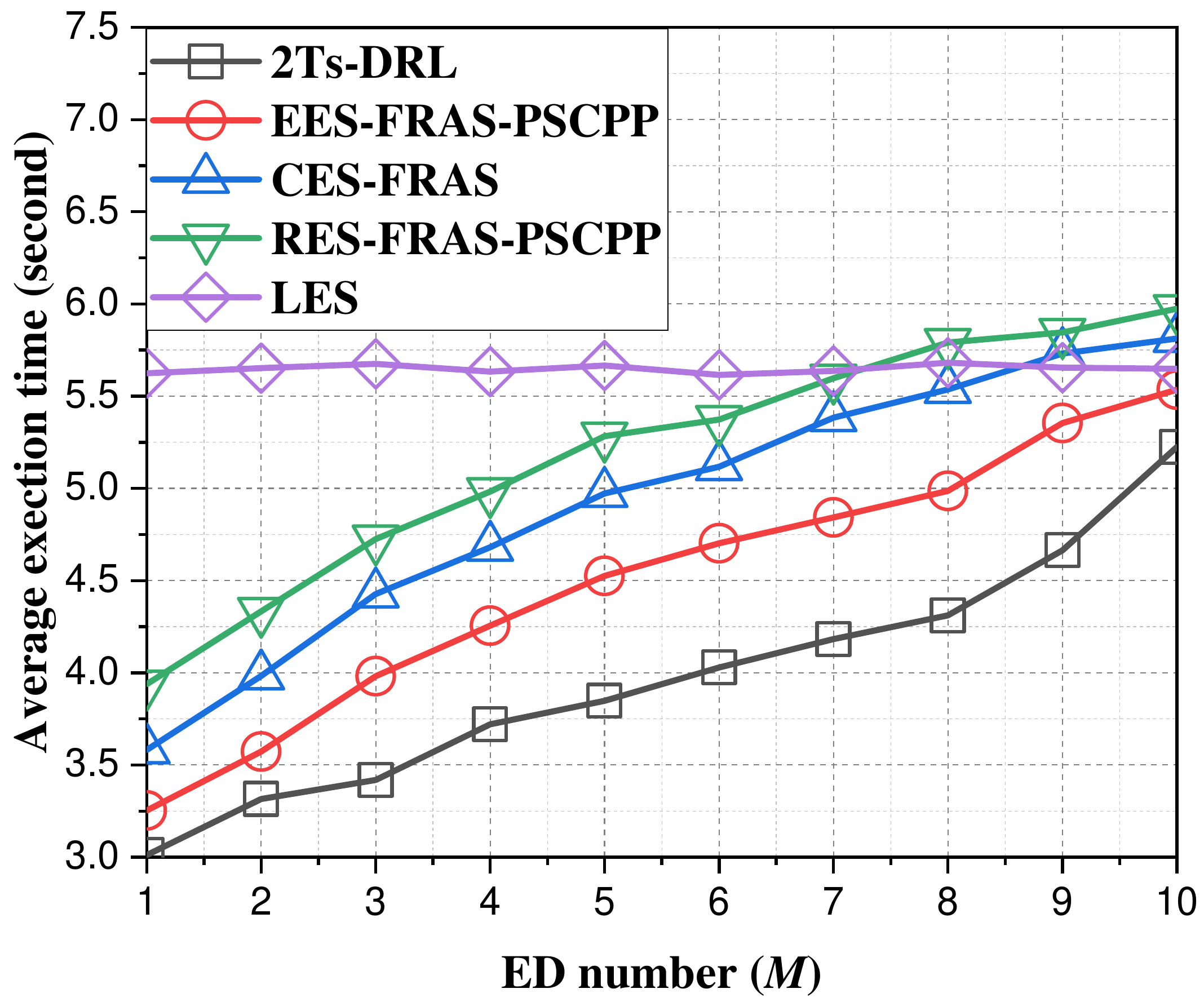}
    \caption{Average task execution time vs. ED density $M$.}
    \label{fig:density}
\end{figure}
Fig.~\ref{fig:density} illustrates the average task execution time for different ED density $M$ to evaluate the performance of our proposal in UDEC scenario.
First of all, our proposal \textit{2Ts-DRL} outperforms the related benchmark policies in term of task execution time.
Note that the LES has stable task execution time when the density grows from 1 to 10, since LES performs local execution, has no additional transmission delay.
The execution time of RES-FRAS-PSCPP, CES-FRAS, EES-FRAS-PSCPP and our proposed \textit{2Ts-DRL} increase when the ED density grows.
Because all the EDs within the same SCceNB coverage share the limited computation and communication resources.
Thus, the performance of the task execution time declines as the number of ED increases.
However, the problem can be solved by deploying more SCceNBs at network edge, especially in some hot-spots areas.
For our \textit{2Ts-DRL}, the average task execution time increases dramatically when $M>8$.
The reason is that each SCceNB has a limited computation capacity, can serve up to 8 EDs in our simulation.
Thus, some of EDs have to perform LES when $M>8$.

\begin{figure}[t!]
    \centering
    \includegraphics[width=3in]{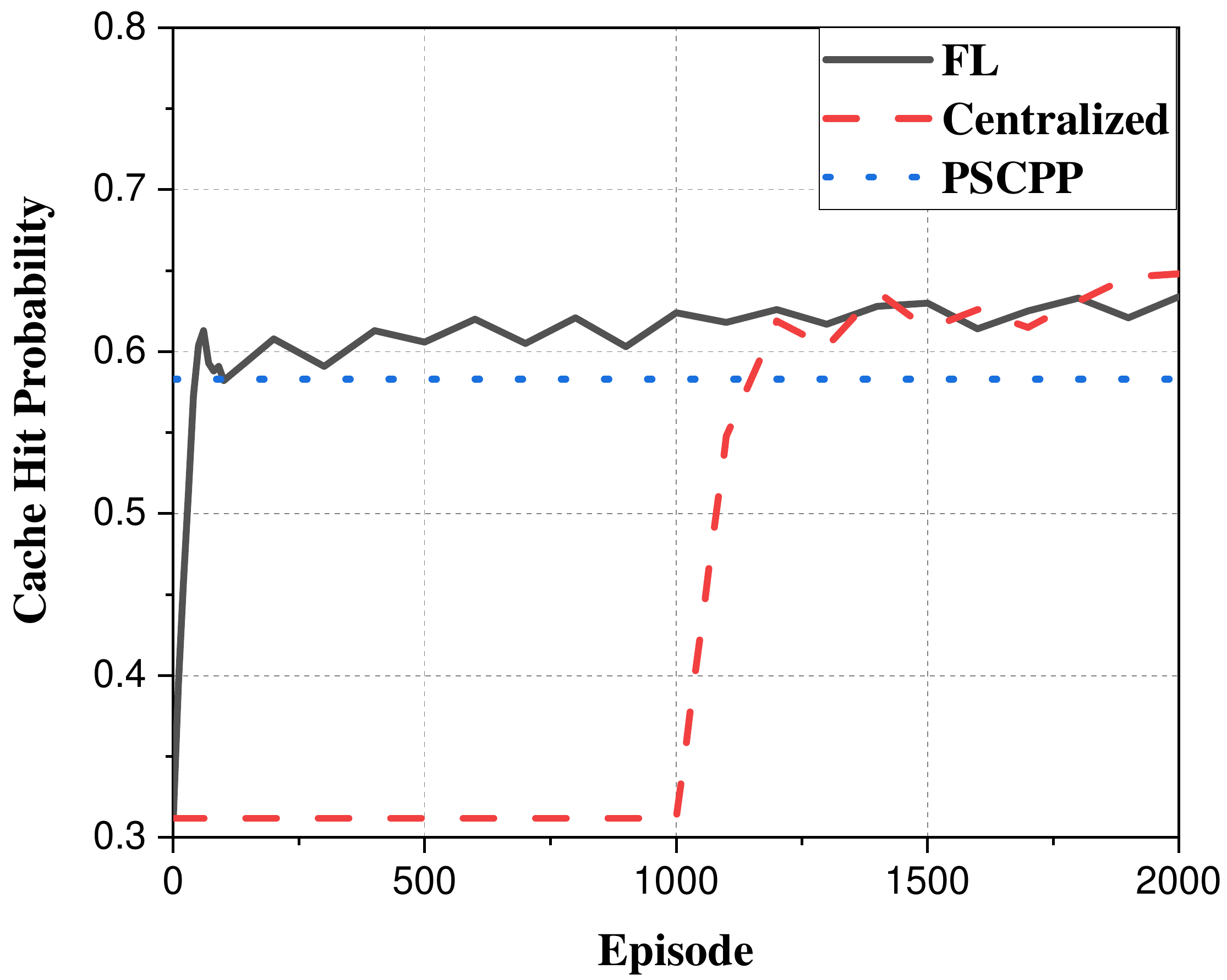}
    \caption{Cache hit probabilities for different service caching placement policies.}
    \label{fig:FL-CHP}
\end{figure}

In this work, we utilize the cache hit probability as the performance metric to evaluate our proposed \textit{2Ts-DRL}.
The metric is the ratio of service cache hits to the number of ED's service requests on the cache.
Fig.~\ref{fig:FL-CHP} reports the cache hit probabilities for different service caching placement policies.
Specifically, FL refers to the proposed federated learning-based model training policy.
Centralized represents a centralized model training policy, which means that the decision agent use the service requests from the last 1000$t$ decision period.
Several observations can be made.
First of all, PSCPP has a fixed cache hit probability, since all the service have predefine popularity indexes, and PSCPP always cache the most popular services.
Then, the Centralized has a fixed cache hit probability from decision period 0 to 1000 when the decision agent collects the service requests from the EDs.
The decision agent begins to train the model at decision period $t=1000$, thus the cache hit probability grows with the number of the episodes increases.
At last, we can find that the cache hit performance of our proposed FL is near close to the results of Centralized.
However, our proposed federated learning-based model training method can protect the data privacy (e.g., most used type of services) for the EDs.

\begin{figure}[t!]
    \centering
    \includegraphics[width=3in]{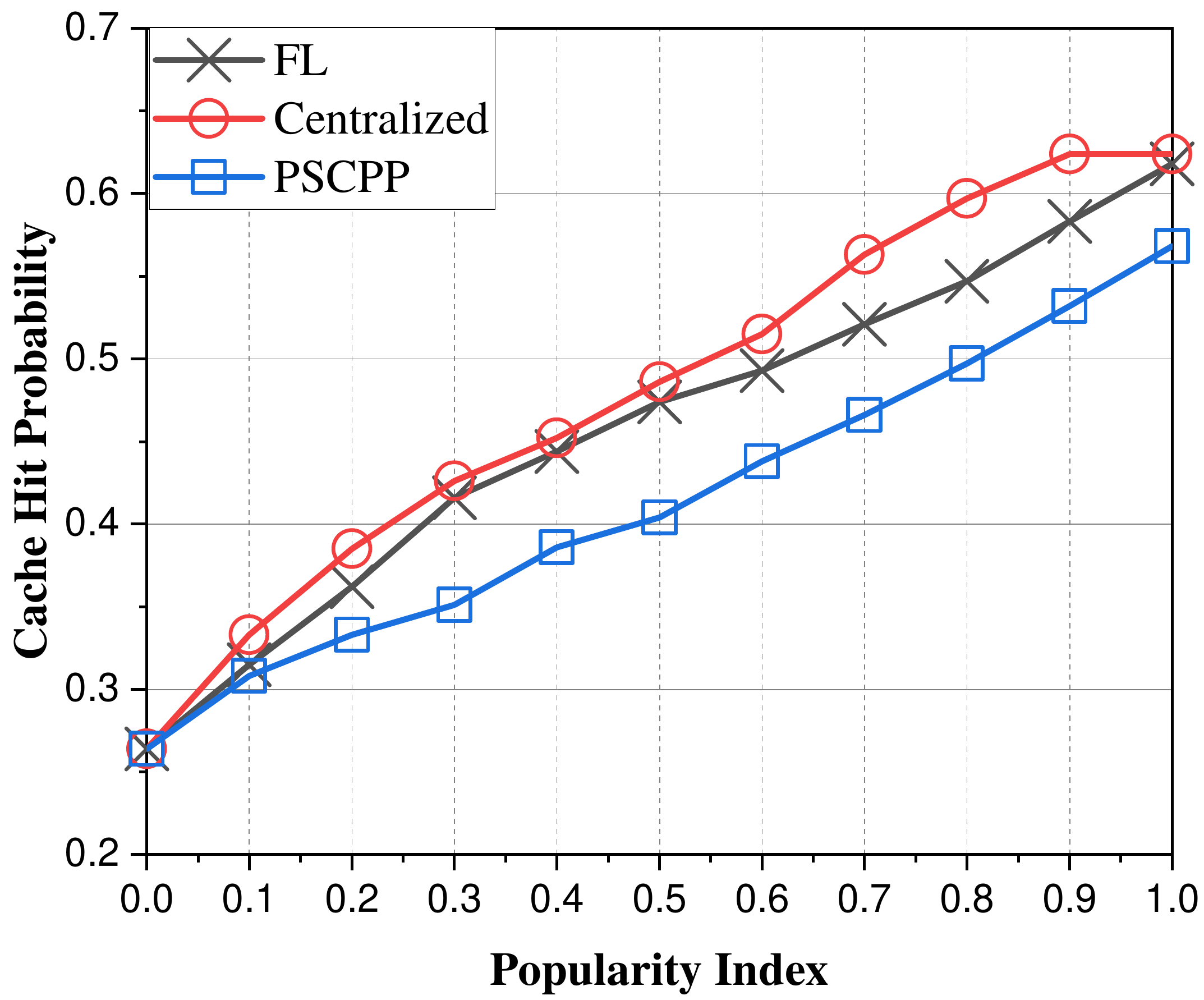}
    \caption{Cache hit probabilities vs. popularity index $\delta$.}
    \label{fig:CHP-INDEX}
\end{figure}

Fig.~\ref{fig:CHP-INDEX} reports the cache hit probabilities of different service caching placement policies under different popularity index $\delta$.
All the policies have the similar cache hit probabilities when $\delta=0$, since the popularity profile is uniform over all the services if $\delta=0$.
The cache hit probability grows with $\delta$ increases for the three policies, and DRL based training policies (both centralized and distributed) always outperform PSCPP in cache hit probability.

\section{Extension and future works}{\label{sec:extension}}
In this section, we will discuss some potential and interesting research directions to extend this work.
\subsection{Blockchain-enabled I-UDEC}
In recent years, blockchain, as the underlying technology of cryptocurrencies, has been adopted in various applications, and attracted much attention.
Note that our proposed I-UDEC still faces some challenges, such as decentralized resource management and security.
Integrating blockchain into I-UDEC is synergistically beneficial for the following reasons.
First, security and privacy are significant challenges to I-UDEC due to the interplay of heterogeneous edge nodes (i.e., SCceNBs and edge devices) and the attack vulnerability of the edge nodes.
In this work, we propose a federated learning-based \textit{2Ts-DRL} algorithm for the I-UDEC to guarantee the privacy of edge devices.
Note that the I-UDEC still face the security challenges, such as DDoS attack and packet saturating.
It makes sense to integrate blockchain into I-UDEC, since the integration can i) replace the expensive key management for multiple communication protocols,
ii) enable easy access for the maintenance of the distributed ultra-dense edge SCceNBs of I-UDEC and
iii) provide efficient monitoring in the control plane to prevent malicious behaviors.
Second, in order to tackle the challenge of resource heterogeneity of I-UDEC network, a distributed resource management scheme is required.
It is possible to build a distributed control at massive SCceNBs by leveraging blockchain, since blockchain has the feature of decentralized control.
The resource management involves edge resource borrowing and lending, its pricing based optimization algorithms will play a key role in resource overhead reduction.
Specifically, smart contract of blockchain enables the use of edge resources (i.e., computation, communication and storage resources) on demand.
Because smart contract can automatically run on-demand resource algorithm for the computation offloading requirements.
Last but not least, blockchain systems are hungry for hardware resource (e.g., computing power), and I-UDEC has rich network resources at network edge.
Thus, the deployment of the blockchain in the I-UDEC is mutually beneficial.
The above research directions will be concerned in our future work.
\subsection{Personalized federated learning-enhanced \textit{2Ts-DRL}}
Note that in practical scenarios, SCceNBs in different areas may have different service caching placement state (i.e., service popularity distribution).
As a result, it is hard to aggregates the local models with different shapes through traditional FL.
To this end, we can use personalized federated learning (PFL)~\cite{PFL,9090366} to capture the characteristics of local service popularity and local offloading requirements.
In PFL, a global model is first trained by standard FL, then, each SCceNB will train a personalized model based on the global model information and its own personal information.
Thus, each SCceNB has a personalized service caching model for different EDs tailored to their computation offloading requirements.
This kind of heterogeneous service caching placement scenario will be studied in our future work.

\section{Conclusions}{\label{sec:conclusion}}
In this article, a joint computation offloading, resource allocation and service caching placement problem is studied for ultra-dense edge computing networks
First, we introduced an intelligent ultra-dense edge computing (I-UDEC) framework in 5G ultra-dense network environments, in order to formulate the heterogeneous network resources and hybrid computation offloading pattern of UDEC.
Then, in order to minimize the task execution time and network resource usage, we optimize the application partitioning, resource allocation and service caching placement in two different timescale.
To this end, we present a two-timescale deep reinforcement learning (\textit{2Ts-DRL}) approach to jointly optimize the above issues for the I-UDEC.
The bottom tier of the \textit{2Ts-DRL} outputs delay sensitive decisions in a fast timescale, whereas the top tier outputs delay insensitive decision in a slow timescale.
Last but not least, we use a FL-based distributed model training method to train the \textit{2Ts-DRL} model, in order to protect edge users' sensitive service request information.
Experimental results corroborate the effectiveness of both the \textit{2Ts-DRL} and FL in the I-UDEC framework.

%\section*{Acknowledgment}
%
%This work was supported in part by the National Key Research and Development Program of China under grant (No.2017YFB1001703); the National Science Foundation of China (No. 62002397 and No. 61972432);
%the Program for Guangdong Introducing Innovative and Entrepreneurial Teams (No.2017ZT07X355);
%the Pearl River Talent Recruitment Program (No.2017GC010465);
%the Science and Technology Program of Guangzhou under Grant 202007040006;
%the Guangdong Special Support Program under Grant 2017TX04X148.

% or


\ifCLASSOPTIONcaptionsoff
  \newpage
\fi



%\bibliographystyle{IEEEtran}
%\begin{thebibliography}{1}


%\end{thebibliography}
%\section*{Acknowledgment}
%This work was supported by the National Science Foundation of China (No. U1711265), the Fundamental Research Funds for the Central Universities (No.17lgjc40), and the Program for Guangdong Introducing Innovative and Enterpreneurial Teams (No.2017ZT07X355), and the FUI PODIUM project (Grant no. 10703).

\bibliographystyle{IEEEtran}
\bibliography{IEEEabrv,bibThesis}
\end{document}